\documentclass[trackchanges,twocolumn]{aastex7}

\usepackage{colortbl}
\accepted{to Journal of High Energy Astrophysics, 2026}

\begin{document}
\begin{flushright} 
\end{flushright}
\title{X-ray spectral scaling of black hole mass in quasi-periodic eruptive source RX~J1301.9+2747} 

\author[orcid=0000-0000-0000-0001,sname='Lev']{Lev Titarchuk}
\affiliation{INAF-IAPS Via Fosso Del Cavaliere 100, 00133, Rome, Italy}
\email{lev.titarchuk@inaf.it}
\affiliation{251 Rue Bernard Blier Tartugui 84400 GARGAS France}

\author[orcid=0000-0001-5129-1491,gname=Seifina, sname='Elena']{Elena Seifina} 
\affiliation{Lomonosov Moscow State University/Sternberg Astronomical Institute,
Universitetsky Prospect 13, Moscow, 119992, Russia \email{seif@sai.msu.ru}}
\email[show]{seif@sai.msu.ru}

\author[gname=Vadim,sname=Arefiev]{Vadim Arefiev}
\affiliation{Space Research Institute of the Russian Academy of Sciences, Profsoyuznaya 84/32, Moscow, 117997, Russia}
\email{arefiev@iki.rssi.ru}

\author[gname=Diana,sname=Makhinya]{Diana Makhinya}
\affiliation{Lomonosov Moscow State University Branch in the City of Sarov, Parkovaya 8, Sarov, Russia}
\email{dina185964@gmail.com}

\author[gname=Alexei,sname=Le]{Alexei Le}
\affiliation{University of Chicago, 5801 S Ellis Ave, Chicago, IL 60637, USA}
\email{alexeile@uchicago.edu}




\begin{abstract}
 We present a spectral and timing analysis of X-ray emission from the actively accreting supermassive black hole (SMBH) in RX~J1301.9+2747 with prominent quasi-periodic eruptions (QPEs) using {\it ASCA,   ROSAT,  XMM-Newton}, and {\it NICER} observations. {\it NICER} detected renewed QPE activity from this source in 2024. 
We examined in detail  the X-ray spectrum formation in this source during QPE outbursts  applying the first-principal model, Bulk Motion Comptonization.  We find an increase of the photon index $\Gamma$ with  a mass accretion rate $\dot M$,  the index saturation at $\Gamma_{sat}=2.7\pm 0.1$ at high $\dot M$  value during outburst peaks. 
Based on the derived $\Gamma-\dot M$ correlation, we apply the scaling technique for the first time to measure the black hole mass in RX~J1301.9+2747.
 The SMBH mass in  this source was estimated to be between $4\times 10^5$ M$_ {\odot}$ and   $7\times 10^6$ M$_ {\odot}$ for the inclination range from 0$^{\circ}$ to 70$^{\circ}$,  which is adopted as a plausible range based on the AGN unified model assumption,  applying   HLX--1~ESO 243--49, OJ~287, and 1H~0707--495 as reference scaling sources. We also estimate  a BH mass in RX~J1301.9+2747  applying a timing analysis, in which 
the Compton cloud size to be $\sim 10^{11}$ cm, estimated from  the quasi-periodic oscillation frequency, $\nu\sim 10^{-4}$ Hz,  indicates SMBH mass in  RX~J1301.9+2747 is of  order of $10^6$ M$_{\odot}$.  Overall, the BH mass obtained by the scaling method in this source is consistent with previous measurements using other methods.
\end{abstract}

\keywords{X-rays: galaxies: active -- galaxies:  nuclei -- quasars: general: quasars: super massive black holes: individuals: RX J1301.9+2747
}




\section{Introduction}

Recently, interest has increased in the quasi-periodic variability of soft X-ray emission emanating from galactic nuclei. These quasi-periodic flares 
 are characterized by relatively high intensity and typically repeat every few hours or weeks. Typical timescales for such flares indicate small, compact, and high-energy, non-periodic, or ``quasi-periodic'' outbursts. They differ from other transient phenomena in their high amplitude (up to $L_X\sim 10^{43}$ erg/s) and short, constant timescales \citep[see][]{Miniutti19,Arcodia21,Chakraborty21,Arcodia22,Arcodia24,Chakraborty24,Hernandez-Garcia25}. Surprisingly, they are often detected in relatively quiet galaxies (between such outbursts, when quiescent, the galactic nucleus exhibits remarkably stable and soft X-ray emission), signaling a new, transient type of activity. These extreme events are presumably associated with supermassive black holes (SMBHs) located at the centers of corresponding galaxies and are characterized by rapid increases and decreases in luminosity. Such variability type in an active galactic nucleus (AGN) was first discovered/identified in X-ray emission from the Seyfert~2 galaxy GSN~069 in December 2018 using {\it XMM-Newton} \citep{Miniutti19}. GSN~069 exhibited peculiar high-amplitude X-ray eruptions on short time scales with an interval of 29.8 ks. Subsequent observations of this source with {\it XMM-Newton} and {\it Chandra} revealed numerous ultra-soft X-ray outbursts  \citep{Miniutti19}. 
The outburst profile was close to symmetric with similar rise and decay times of about 1.8 ks, and the total event duration was $\sim$4.5 ks for each outburst. 
\cite{Miniutti19} made up  the term X-ray quasiperiodic eruptions (QPEs) to distinguish these observed phenomena from the softer, quasi-sinusoidal modulation of standard quasi-periodic oscillations (QPOs) often observed in X-ray binaries \citep{van der Klis89} and several supermassive accreting BHs \citep{Lin13,Gierlinski08}. 

During QPOs, the X-ray spectrum was shown to oscillate between cold (around 50 -- 80 eV at the beginning and end of the outburst) and warm (120 -- 200 eV at the peak of the outburst) phases. These phases were subsequently termed {\it quiescent} (QS) and {\it eruptive} (ES) states \citep[see e.g.,][]{Miniutti23,Giustini20,Giustini24}. It has been suggested that the variable thermal emission may be physically related to the Comptonization of low-energy disk photons in the warm, optically thick  corona or, for example, to emission from different disk regions that are activated at different times during the cycle. \cite{Miniutti19} also noted similar properties of GSN~069 and another AGN, RX J1301.9+274715 (RXJ1301 hereafter), in terms of repeated soft X-ray outbursts \citep{Sun13}. There are currently nine such unique objects, but the mechanism underlying their variability is still not well understood.

%
%
\begin{table}
\centering
\caption{Basic parameters of RX~J1301.9+2747.}
\label{tab:parameters_rx}
 \begin{tabular}{llrcl}
 \hline\hline                        
  Parameter               &     Value    \\
      \hline
Source class                                     & Seyfert$^{~(a)}$                   \\
Mass of a primary BH, M$_{\odot}$ & $\sim (0.8-2.8)\times 10^{6~(b,c)}$,\\
                                                   & $\sim$1.26~$\times ~10^{6~(d,e)}$    \\
Mass of a secondary BH, M$_{\odot}$ & $\sim 1.4^{~(f)}$ -- 8.8~$\times ~10^{2~(e)}$  \\
Distance $D$,  Mpc                            & $\sim$ 99.3$^{(d)}$                    \\
Red shift, $z$                                      & 0.02369$\pm$0.00017$^{(c)}$ \\
RA (J2000), $\alpha$       & 13$^h$ 02$^m$ 00.096$^{s~(c,g)}$      \\
Dec (J2000), $\delta$       & +27$^{\circ}$ 46{\tt '} 57.51{\tt "}$^{~(c,g)}$      \\
QPO frequency              & \\
during quiescence,  $\nu_{QPO}$,  Hz       & 3 $\times$ 10$^{-4~(a)}$                       \\
$N_{H;Gal}$, cm$^{-2}$                      & $8.2 \times 10^{19~(c,h)}$                    \\
Alternative name &1WGA J1301.9+2746$^{(i)}$,\\
                         &CGCG 1259.6+2803, \\
                         & CG 0959, FOCA 0357,\\
                         & PGC 044947, Coma-D 61,\\
                         & RX J1301.9+2746,\\
                         & KUG 1259+280, GMP 1681\\
 \hline                                             
\end{tabular}
\begin{flushleft}
$^{(a)}$ \cite{Sun13,Dewangan00}; 
$^{(b)}$ \cite{Shu17};   
$^{(c)}$ \cite{Giustini20,Caldwell+Rose97}; 
$^{(d)}$ \cite{Middleton25};
$^{(e)}$ \cite{Franchini23};
$^{(f)}$ \cite{Zhao22};
$^{(g)}$ \cite{Yang22};
$^{(h)}$ \cite{Kalberla05},
$^{(i)}$ WGA Catalog of ROSAT point Sources (WGACAT). 
\end{flushleft}
\end{table}

A prominent example of a QPE-source is RXJ1301, which is an ultra-soft AGN \citep{Dewangan00} with unusual X-ray variability, characterized by a long-term quiescence state and a short-term outburst (eruptive) state \citep{Giustini24, Giustini20,Song20} with a periodic repetition time scale of 13--20 ks (other source parameters are listed in Table~\ref{tab:parameters_rx}). \cite{Giustini24} analyzed X-ray emission from RXJ1301 in both states using data from XMM-Newton, EXOSAT, ROSAT, and Chandra. They found that QPEs in RXJ1301 do not exhibit the strict regularity observed in the discovery source, GSN~069. In terms of timing characteristics, there is no clear pattern of repetition between QPEs in RXJ1301: the average time between their peaks was about 4~h, but could be as short (1~h) or as long (6~h). However, the spectral characteristics of RXJ1301's QPEs, depending on energy, are very similar to those of GSN~069 and other QPE sources. \cite{Giustini24} showed that during their evolution, X-ray QPEs in RXJ1301 follow a hysteretic pattern in the temperature-luminosity plane, with a hotter  rise than a decay. The quiescent emission of RXJ1301 turned out to be more complex than that of GSN~069, since in addition to thermal emission from the accretion disk, it also requires a component similar to a soft X-ray excess. Its long-term variations in the quiescent X-ray flux have low amplitude and are not strictly monotonic, with an overall decline over 22 years. To date, XMM-Newton has detected 34 QPEs \citep{Giustini24}. 

Various models have been proposed to explain the above observed QPE-events, including the self-lensing supermassive BHs (SMBHs) \citep[][hereaftre G24]{Ingram21, Giustini24}.  The ``Swiss cheese'' rotating wind model \citep{Middleton+Ingram15},  the radiative disk instability \citep[e.g.,][]{Raj+Nixon21,Kaur23,Pan23}, and interactions (tidal interactions or disk crossings/punctures) with a lower-mass secondary, including a white dwarf \citep{King23} and a stellar-mass BH \citep{Song20}, often referred  as  extreme mass-ratio in-spiral
(EMRI) systems \citep{Dai10,Dai+Blandford13,Xian21,Krolik+Linial22,Metzger22,Wang22,Zhao22,Franchini23,King23,Linial+Metzger23,Linial+Metzger24,Linial+Sari23,Lu+Quataert23,Tagawa+Haiman23,Yao25, Zhou24} and an accretion disk model with the Lense-Thirring precession that also allows  a super-Eddington wind \citep{Middleton25}. The EMRI scenario assumes the presence of an accretion disk around a SMBH interacting with a stellar-mass object in a close orbit \citep{Dai10,Xian21,Linial+Metzger23}. Within this model, the masses of SMBHs for QPE sources are in the range of $10^5 - 10^7$ M$_{\odot}$ \citep{Wevers22,Arcodia24}. Application of this model to RXJ1301 indicates a primary black hole mass of M$_1 = 10^{6.3}$M$_{\odot}$ and a secondary black hole mass of M$_2 \le$ 100 M$_{\odot}$ \citep{Franchini23}.  In turn, \cite{Shu17} estimated the mass of the primary BH  in RXJ1301 using the UV/X-ray analysis, $M_{BH}\sim (0.8-2.8)\times 10^6$ M$_{\odot}$  \citep[see also][]{Giustini20}.

To test the EMRI model proposed for interpreting the X-ray variability of RXJ1301, we aimed to estimate a BH mass  in RXJ1301 using an alternative BH weighting method, the X-ray spectral scaling method \citep[][hereinafter -- ST09]{st09}.  This method was developed using correlation scaling between the photon index $\Gamma$ and the normalization  of the X-ray spectrum (or the mass accretion rate $\dot M$), measured with the {\it Bulk Motion Comptonization}  (BMC) model and observed in many galactic binaries with BHs during transitions between QS and ES states.  A condition for this method applicability is that the source exhibits a change in spectral states accompanied by a characteristic behavior of the index $\Gamma$ during an outburst: $\Gamma$ increases monotonically with $\dot M$ during the transition QS$\to$ES  and reaches a maximum constant level for a given source at high values of $\dot M$. Then $\Gamma$ decreases monotonically during the transition ES$\to$QS as the outburst decays. The saturation of $\Gamma$ (the so-called $\Gamma$-saturation phase, at which $\Gamma$ reaches a constant value in our parameter space) during the outburst is a specific signature that the given object contains a BH \citep{tz98}. In fact, the $\Gamma$-saturation phase is best explained by  an accretion flow converging to the BH event horizon [see, e.g., the numerical simulation results obtained by \cite{laurent_titarchuk07,LT99}]. Then, assuming that the BH accretion disk luminosity is directly proportional to $\dot M$ (and hence to the mass of the central BH) and inversely proportional to the square of the distance to the source, we can estimate the mass of a black hole with a known distance to it (and an unknown mass) by comparing its track $\Gamma-\dot M$ with the track $\Gamma-\dot M$ of a black hole source for which the mass and distance are known (for more details on the scaling method, see \cite{tss10,ST10}, and ST09). 

%
%

\begin{figure*}
\centering
\includegraphics[width=18.1cm]{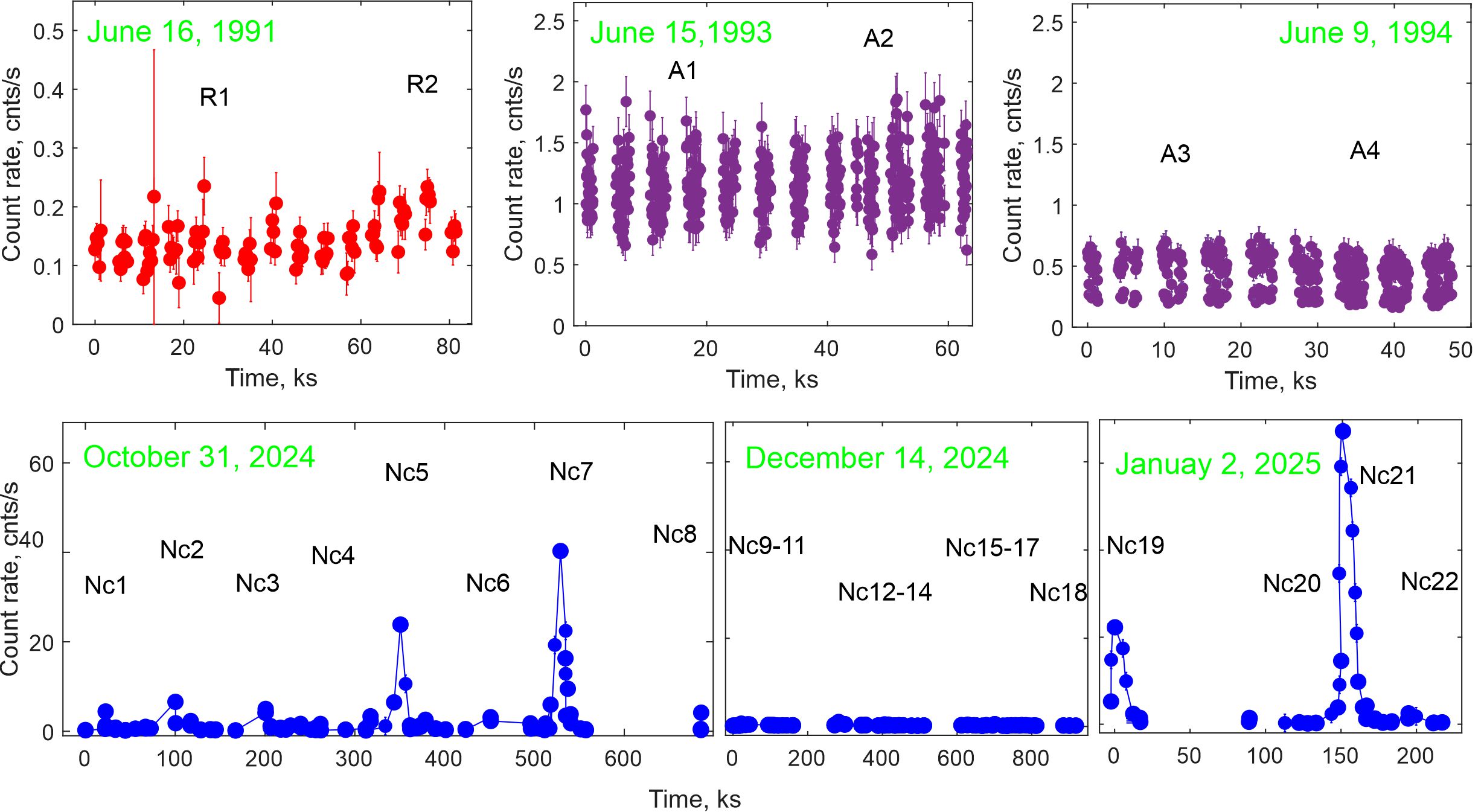}
\caption{X-ray binned light curves of RXJ1301 in the 0.2--2 keV energy range obtained by  {\it  ROSAT} (red points, top panels), {\it  ASCA} (violet points, top panels),  and by {\it NICER} (blue points, middle panels) with timebin of 100 s.  The UT start date of the light curve is indicated at the top of each panel  in green. The observation epochs by different missions are indicated along the corresponding light curves in black, as listed in Table~\ref{tab:table_Suzaku+ASCA_SAX}.
}
\label{ev_rx_xmm}
\end{figure*}

In Sect.~\ref{data} we describe the observations and data processing. In Sect.~\ref{results} we present the light curve analysis (Sect.~\ref{lc}), spectral analysis (Sect.~\ref{spectral analysis}), and the BH mass estimate in RXJ1301 (Sect.~\ref{mass_estimate}). A discussion of the results obtained and their implications is presented in Sect.~\ref{discussion}. We make our final conclusions in Sect.~\ref{summary}.

%
%
\begin{table*}
 \centering 
 \caption{Details of  {\it ROSAT},  {\it ASCA}, {\it XMM-Newton}, and {\it NICER}  observations of RX~J1301.9+2747. 
}
   \label{tab:table_Suzaku+ASCA_SAX}
\resizebox{1.\textwidth}{!}{%
 \begin{tabular}{llllllr}
 \hline\hline  
Epoch & Mission & Obs. ID& Start time (UT)  & End time (UT) & MJD interval & Source\\
 & & & && & state\\
      \hline
R1 & ROSAT     & RP800006N00  & 1991 June 16 22:46:38 & 1991 June 17 22:44:33 &48423.9--48424.9&QS\\ 
R2 & ROSAT     & RP800005N00  & 1991 June 17 22:45:02 & 1991 June 18 22:43:10 &48424.9--48425.9&QS\\ 
\hline
A1 & ASCA           & 80019000  & 1993 June 15 09:25:38 & 1993 June 15 18:32:28 &49153.3-49153.8&QS\\
A2 & ASCA           & 80020000  & 1993 June 15 18:57:02 & 1993 June 16 03:00:00 &49153.8-49154.1&QS\\
A3 & ASCA           & 82019000  & 1994 June 9 23:50:15 & 1994 June 10 07:49:49 &49512.9-49513.3&QS\\
A4 & ASCA           & 82020000  & 1994 June 10 07:50:23 & 1994 June 10 14:10:37 &49513.3-49513.6&QS\\
\hline
X1 & {\it XMM-Newton}    & 0124710801  & 2000 Dec 10 20:13:20 & 2000 Dec 11 02:24:00  &51888.8--51889.1&QS,ES\\ 
X2 & {\it XMM-Newton}    & 0851180501  & 2019 May 30 20:42:24 & 2019 May 31 10:09:04  &58633.8--58634.4&QS,ES\\
X3 & {\it XMM-Newton}    & 0864560101  & 2020 Jul 11 18:14:03 & 2020 Jul 13 06:49:36 &59041.7--59043.2&QS,ES\\ 
X4 & {\it XMM-Newton}    & 0904640201 & 2022 Jun 17 19:47:19 & 2022 Jun 19 06:50:39 &59747.8--59749.3&QS,ES\\
X5 & {\it XMM-Newton}    & 0904640301  & 2022 Jun 19 09:40:51 & 2022 Jun 21 06:42:31 &59749.4--59751.3&QS,ES\\
\hline
Nc1 & {\it NICER}  & 7202900101  & 2024 Oct 31 04:57:52 & 2024 Oct 31 23:39:52 &60614.2--60614.9&QS\\
Nc2 & {\it NICER}  & 7202900102  & 2024 Nov 1 01:04:48 & 2024 Nov 1 21:20:02 &60614.2--60615.9&QS\\
Nc3 & {\it NICER}  & 7202900103  & 2024 Nov 2 03:23:01 & 2024 Nov 2 23:38:41 &60616.1--60616.9&QS\\
Nc4 & {\it NICER}  & 7202900104  & 2024 Nov 3 02:40:37 & 2024 Nov 3 21:14:59 &60617.1--60617.9&QS\\
Nc5 & {\it NICER}  & 7202900105  & 2024 Nov 4 04:53:12 & 2024 Nov 4 20:27:39 &60618.2--60618.8&QS,ES\\
Nc6 & {\it NICER}  & 7202900106  & 2024 Nov 5 02:32:50 & 2024 Nov 5 22:46:00 &60619.1--60619.9&QS\\
Nc7 & {\it NICER}  & 7202900107  & 2024 Nov 6 03:18:25 & 2024 Nov 6 15:47:20 &60620.1--60620.6&QS,ES\\
Nc8 & {\it NICER}  & 7202900108  & 2024 Nov 8 03:16:58 & 2024 Nov 8 03:28:03 &60622.13--60622.18&QS\\
Nc9 & {\it NICER}  & 7202900109  & 2024 Dec 14 11:16:32 & 2024 Dec 14 17:42:10 &60658.4--60658.7&QS\\
Nc10 & {\it NICER}  & 7202900110  & 2024 Dec 15 04:19:29 & 2024 Dec 15 23:00:20 &60659.1--60659.9&QS\\
Nc11 & {\it NICER}  & 7202900111  & 2024 Dec 16 06:42:11 & 2024 Dec 16 22:19:20 &60660.0--60660.9&QS\\
Nc12 & {\it NICER}  & 7202900112  & 2024 Dec 17 02:46:42 & 2024 Dec 17 23:04:20 &60661.1--60661.9&QS\\
Nc13 & {\it NICER}  & 7202900113  & 2024 Dec 18 01:55:25 & 2024 Dec 18 23:49:40 &60662.0--60662.9&QS\\
Nc14 & {\it NICER}  & 7202900114  & 2024 Dec 19 02:46:52 & 2024 Dec 19 21:26:33 &60663.1--60663.9&QS\\
Nc15 & {\it NICER}  & 7202900115  & 2024 Dec 21 01:11:19 & 2024 Dec 21 19:51:20 &60665.0--60665.8&QS\\
Nc16 & {\it NICER}  & 7202900116  & 2024 Dec 22 00:21:40 & 2024 Dec 22 22:07:40 &60666.0--60666.9&QS\\
Nc17 & {\it NICER}  & 7202900117  & 2024 Dec 23 01:08:27 & 2024 Dec 23 08:58:07 &60667.0--60667.4&QS\\
Nc18 & {\it NICER}  & 7202900118  & 2024 Dec 24 05:00:19 & 2024 Dec 24 14:24:40 &60668.2--60668.6&ES\\
Nc19 & {\it NICER}  & 7202900119  & 2025 Jan 2 12:02:06 & 2025 Jan 2 16:47:00 &60677.50--60677.69&QS,ES\\
Nc20 & {\it NICER}  & 7202900120  & 2025 Jan 3 12:46:45 & 2025 Jan 3 23:38:40 &60678.53--60678.98&QS\\
Nc21 & {\it NICER}  & 7202900121  & 2025 Jan 4 01:04:06 & 2025 Jan 4 22:49:29 &60679.04--60679.95&QS,ES\\
Nc22 & {\it NICER}  & 7202900122  & 2025 Jan 5 00:16:49 & 2025 Jan 5 200:22:20 &60680.01--60680.02&QS\\
      \hline
      \end{tabular}
}
\begin{flushleft}
Columns 1 -- 7 denote Epoch, Mission Name, Observation ID,  Start time (UT), End time (UT), MJD interval, respectively. See the text for details.
\end{flushleft}
\end{table*}

%
\begin{figure}
\centering
\includegraphics[scale=1.4,angle=0]{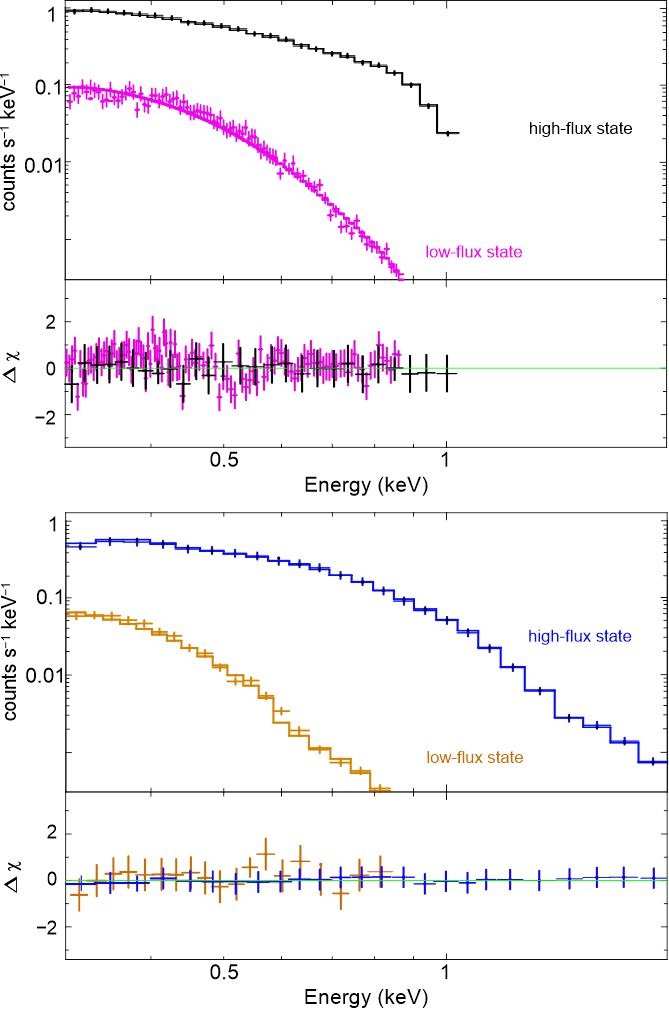}
\caption{XMM-Newton/PN flux-resolved spectra of RXJ1301 for observation $X1$ high-flux (black) and low-flux (pink) in the top panel, as well as $X2$ high-flux (blue) and low-flux (orange) in the bottom panel using model {\tt Tbabs*BMC}. Bottom panels indicate $\Delta \chi$ versus photon energy in keV. 
} 
\label{sp_compar_xmm}
\end{figure}

%
%

\section{Observations and data reduction \label{data}}

We used observations of RXJ1301 with 
 {\it ROSAT} (1991), {\it ASCA} (1993 -- 1994), {\it XMM-Newton} (2000)  and with {\it NICER} (2024). This data is extracted from the HEASARC archives (see Table \ref{tab:table_Suzaku+ASCA_SAX}).

\subsection{{\it ASCA data} \label{asca data}}
{\it ASCA} observed RXJ1301 on  June 15 1993 and June 9 -- 10, 1994 
(see Table~\ref{tab:table_Suzaku+ASCA_SAX}, which  summarized the ID, start time, end time, and the MJD interval). One can see a description of the {\it ASCA}  data  by \cite{Tanaka94}.  The solid imaging spectrometers (SIS)  operated in Faint CCD-2 mode. The {\it ASCA} data were screened using the ftool ascascreen and the standard screening criteria. The spectrum for the source were extracted using spatial regions with a diameter of 4${\tt '}$ (for SISs) and 6${\tt '}$ (for GISs) centered on the nominal position of the source,
while background was extracted from source-free regions of comparable size away from the source. The background-subtracted spectrum data were re-binned to provide at least 20 counts per spectral bin to validate the use of the $\chi^2$-statistic. The SIS and GIS data were fitted using {\tt XSPEC} in the energy ranges of 0.6 -- 2 keV and 0.8 -- 2  keV, where the spectral responses are well-known.  

\subsection{\it ROSAT data \label{rosat data}}

The field of view (FOV) of the source RXJ1301 was observed by the Position-Sensitive Proportional Counter (PSPC) on board the {\it ROSAT} satellite \citep{Truemper83} on June 16--18, 1991 with an exposure of $\sim$44 ks, in total (see also \cite{Dewangan00}). To remind a reader, {\it ROSAT}/PSPC [0.2--2.5 keV] has a 2$^{\circ}$ FOV. It has a spatial resolution of 25{\tt "} 
and an energy resolution of $\Delta E/E = 0.43(E/0.93)^{-0.5}$. Due to the effectively suppressed particle background of PSPC, the overall background is very low \citep{Snowden92}. 
The {\it ROSAT}/PSPC  data of RXJ1301 are available through the HEAsoft archives (HEASARC 2014) and processed using the ftools software package (see Fig. \ref{ev_rx_xmm}. 
Using these data in the 0.1--2.1 keV energy range we studied X-ray characteristics of RXJ1301 in 1991 (Table~\ref{tab:table_Suzaku+ASCA_SAX}). The background was estimated using a nearby source-free annual region of width 60{\tt"} centered at RXJ1301, 2.5{\tt'} from RXJ1301.

\subsection{{\it XMM-Newton} data\label{xmm-n data}}

We used observations of RXJ1301 applying  the European Photon Camera (EPIC) onboard the XMM-Newton satellite \citep{Struder01} in December 10--11, 2000 with an exposure time of 27 ks (see Table~\ref{tab:table_Suzaku+ASCA_SAX})  in the  Full Frame mode applying the EPIC/MOS cameras and in the extended Full Frame mode with the PN camera. The XMM-Newton data were processed according to standard procedures using Science Analysis Software (SAS) software (version 18.0.0)\footnote{https://www.cosmos.esa.int/web/xmm-newton/sas-threads}. Source events were extracted from a circular region with a 40{\tt"} radius centered at the source position (Table~\ref{tab:parameters_rx}), while background events were from source-free regions on the same chip using four circular regions identical to the source region. The SAS tool {\tt epiclccorr}  was used to correct for instrumental effects. We did not take into account the photon pile-up effect, as it is negligible at rest due to low count rates.

\subsection{{\it NICER} data\label{nicer data}}
NICER~\citep{Gendreau12} observed source RXJ1301 from October 31 to December 21, 2024 is used together with the latest CALDB to process the data from the observations. 
We perform standard calibration and screening in the NICER archive data with {\tt nicerl2} task. We then use the {\tt nicerl3-spect} task to generate source and background spectra and corresponding response files  in the energy range 0.5--2 keV for each observation. The background-subtracted spectra are rebinned to have a minimum of 20 counts per energy bin for spectral modeling. The publicly available software {\tt stingray}\footnote{https://docs.stingray.science/en/stable/index.html} v2.1 is used to generate the light curve from the original event files. Throughout the paper, uncertainties are presented at the 90\% confidence level. Note that \cite{Sun13} previously analyzed X-ray images of RXJ1301 using Chandra data and showed minimal influence from other nearby sources (e.g., peripheral sources of the Coma cluster) in the vicinity of RXJ1301. Therefore, 
it can be assumed that 
NICER data is not influenced by other nearby sources. 
Nevertheless, we do not rule out that {\it NICER} data may suffer from some background contamination, and this could potentially affect the derived model parameters in the source low state.

 \section{Results \label{results}}
The main goal of our paper is to estimate the BH mass in RXJ1301 using the spectral scaling method with the {\tt BMC} model. Therefore, we will next fit the {\tt BMC} model to X-ray spectra at different epochs. However, since RXJ1301 exhibits QPEs, the spectral properties of which differ during eruptions (ES states) from those during quiescence (QS states), the source spectral properties may be mixed within a single observation. To avoid mixing these different spectral states, before analyzing the spectra of RXJ1301, we examine light curves with a time resolution of $\sim$100 s to identify and separate QPE episodes. This procedure of identifying QPEs from light curves (Section~\ref{lc}) is essential before moving on to a reliable spectral analysis (Section ~\ref{spectral analysis}).

\subsection{X-ray Light Curve of RX~J1301.9+2747 \label{lc}}

Figure~\ref{ev_rx_xmm} presents the light curves of RXJ1301, demonstrating the source X-ray variability in the soft X-rays with {\it ROSAT} (top left),  {\it ASCA} (top middle and right), {\it NICER} (bottom). Interestingly, in the 1993 and 1994 observations, RXJ1301 did not show significant variability, while in 2024--2025 the source deminstrated significant activity. Thus, in October--November 2024 (MJD 60618 and 60620), we found 
 a number QPE-outbutsts: some medium-amplitude QPE-outbursts, and one high-amplitude QPE-outburst in  January 2025 (MJD 60679.3). 
Note that the count rate scale may differ slightly for light curves of different telescopes; however, we used the variability patterns of the light curve. 
 By a  QPE event, we meant an outburst with the relative count rate variation exceeding 40\% of the average count rate for the quiescent state with a QPE duration of 1.5--2 ks. 
Throughout this paper, to describe the stages of the RXJ1301 outburst, we apply a classification of QPE states in AGNs used in \cite{Miniutti19,Sun13}, 
 based on specific types (patterns) of the X-ray source light curve. 
The behavior of RXJ1301 in 1993--1994 can be attributed to a completely QS state, when the count rate was $\sim$1.25 counts/s on average for a relatively bright state with the maximum count rate of 
 $\sim$1.9 counts/s (1993, $A1$ and $A2$) and $\sim$0.4 counts/s for a fainter state with the maximum count rate of $\sim$0.6 counts/s (1994, $A3$ and $A4$). 
 Between QPE-outbursts, the source exhibits only minor X-ray variability during quiescent periods. Background noise has been subtracted from all light curves presented. In general, the source was well detected in all observations above the background level ($3\sigma$).

Previously, \cite{Dewangan00} presented the ROSAT X-ray light curve for RXJ1301 in the energy range 0.1--1.5 keV \citep[see Fig.~4 in][]{Dewangan00}. Here, we present the 0.1--2.5 keV light curve of RXJ1301 to re-test the source state during June 16--18, 1991 ($R1$ and $R2$), for further spectral analysis (see top left panel in Fig.~\ref{ev_rx_xmm} and Section~\ref{spectral analysis}). It is worth mentioning that EXOSAT count rate measurements from RXJ1301 in the 0.05--2.0 keV range in 1985 yielded values similar to those obtained by RASS in 1990, but 25\% lower than during the observation in 1991 \citep{Giommi91}. Thus, the X-ray intensity from RXJ1303 was low during the 1985 and 1990 observations, after which it increased slightly (by 25\%) in 1991. In 1993 and 1994, it also remained consistently low compared to the highly variable emission of RXJ1301 detected from 2018 onwards (see Figure~1 in \cite{Giustini24} and our Figure~\ref{ev_rx_xmm}).

\subsection{Spectral Analysis \label{spectral analysis}}

\subsubsection{Flux-resolved Spectra \label{flux-resolved-spectra}}
After checking the binned light curves for QPE events 
and identifying source states, we divided these observations into segments with high and low count rates for spectral analysis RXJ1301. This eliminated the problem of mixing different states in each observation. 
%
For the {\it NICER} observations, we performed a procedure of separating into high-count rate ($\ge$2 cnt/s) and low-count rate ($\le$2 cnt/s). It turned out that each segment ($\sim$100--200 s) of the full observation also showed count rate variability relative to the mean level for each observation. Therefore, we grouped the high-count rate and low-count rate segments, then constructed the corresponding spectra for each observation and estimated the variation in spectral parameters over the course of a single observation. For the {\it NICER} data, this procedure did not affect observations with very low count rates or short exposure times. 
For the XMM-Newton data, following previous studies by \cite{Giustini20}, we also divided the data into low-flux and high-flux states using threshold values of $<0.2$ count/s and $>0.4$ count/s applied to the EPIC-pn light curves in the 0.2--2 keV range. 
For the {\it ROSAT} 
 and {\it ASCA} observations, we did not perform this procedure, because RXJ1301 does not exhibit QPE behavior in these observations, where the data point errors are quite large. The high and low count rates are likely just fluctuations 
(see Fig.~\ref{ev_rx_xmm}). 
 A similarly weak scatter of light curve points is also visible in the detailed X-ray light curves of RXJ1301 obtained in the 0.1--0.3 and 0.3--1.5 keV energy bands with {\it ROSAT} (see Fig.~4 in \cite{Dewangan00}). Therefore, we assumed that QPE events in RXJ1301 had not yet begun in 
 either 1991 or 1993--1994. 
Therefore, we performed count rate-based state identification (see Section 3.2) using the Bayesian method developed by \cite{Park06}. 
In addition, we applied the effective area option of the Park code, which includes count rate correction factors in their calculations. 

%
%

\begin{figure}
\includegraphics[scale=0.9,angle=0]{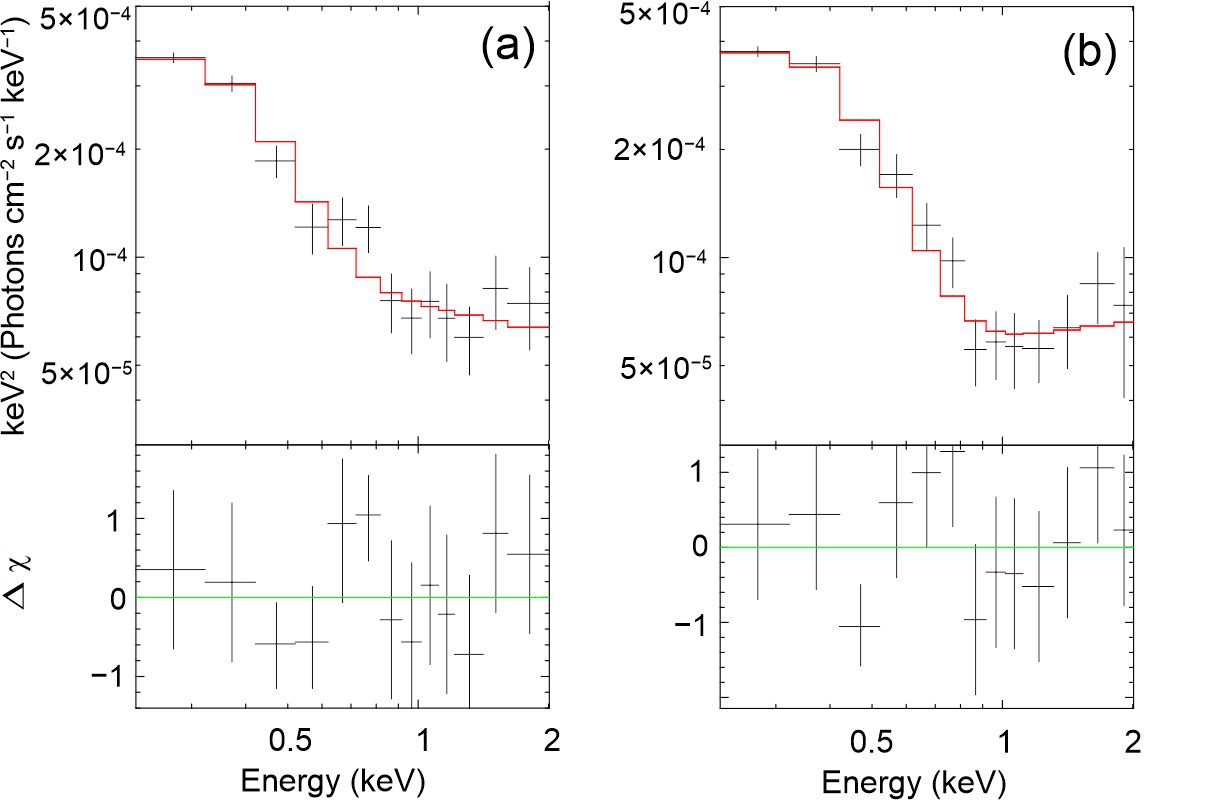}
\caption{{\it ROSAT/PSPC} spectrum of RXJ1301 in units of $E\times F(E)$ for observations $R1$ (a) and $R2$ (b) using model {\tt Tbabs*BMC}. Data are denoted by black crossed and best-fit spectrum indicated by red line. Bottom panels indicate $\Delta\chi$ versus photon energy in keV. See also the best-fit parameters listed in Table \ref{tab:fit_table_1301}.
}
\label{sp_rosat}
\end{figure}

After identifying the QPE episodes in our data, we will fit the corresponding X-ray spectra without mixing states using a spectral model.

\subsubsection{Model Selection \label{model selection}}
\begin{figure}
\includegraphics[scale=0.99,angle=0]{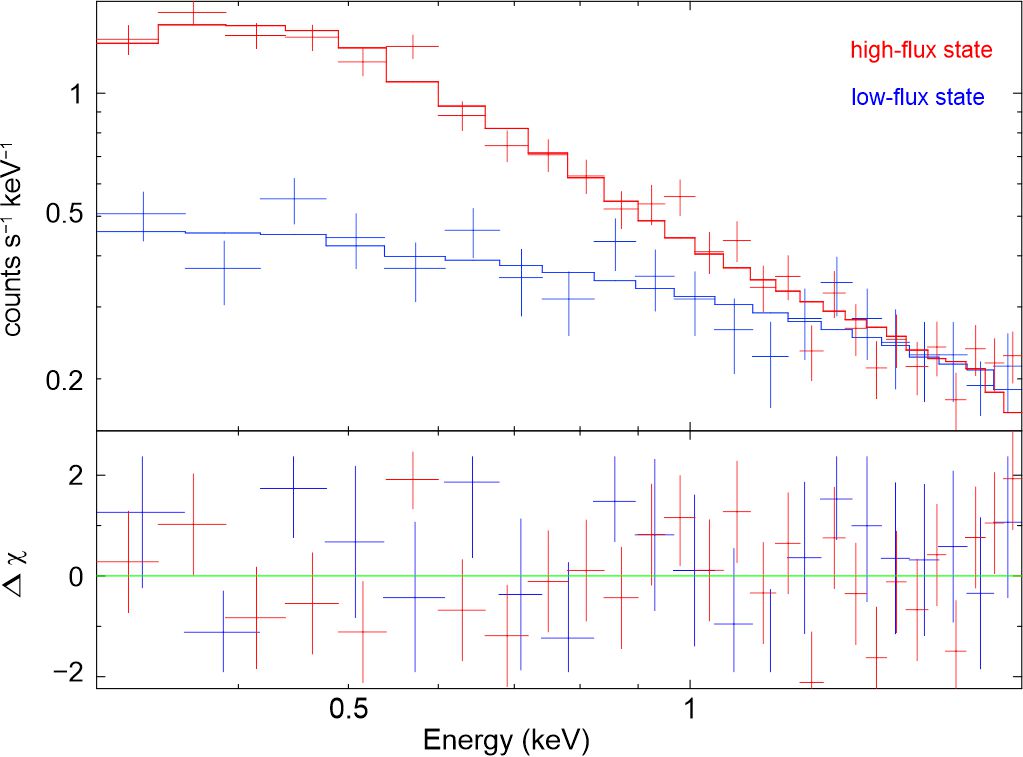}
\caption{Flux-resolved $E*F(E)$ spectra of RX~J1301.9+2747 obtained with {\it NICER} in 2024  observation $N19$ 
with the best-fit modeling using {\tt Tbabs*BMC}. %
The data are denoted by  crosses, while the spectral model is shown by a red (high-flux) and blue (low-flux) histograms for each state.  
Bottom: $\Delta \chi$ for the {\tt BMC} model for the low-flux (blue) and high-flux (red) states versus photon energy in keV. 
}
\label{2_spectra}
\end{figure}

We studied the X-ray spectrum of RXJ1301, taking into account that this AGN lacks a hot corona due to the absence of hard photons. In fact, the X-ray emission from RXJ1301 is supersoft, as its photon energy is limited to 2 keV (see for example Figs.~\ref{sp_compar_xmm}, \ref{sp_rosat}, \ref{2_spectra}, and \ref{8_spectra_rx}). In this case, an additional component besides the blackbody (BB) emission could be a warm corona (similar to a hot corona, but with a lower plasma temperature, well described by the {\tt CompTT}\footnote{https://heasarc.gsfc.nasa.gov/docs/software/xspec/manual/node159.html} model \citep{Titarchuk94}) and/or an accretion disk (the emission of which is well described by the {\tt BMC}\footnote{https://heasarc.gsfc.nasa.gov/xanadu/xspec/manual/node142.html} model \citep{TMK97,tz98}). 
These latter models are believed to describe two possible Comptonization mechanisms and are physically distinct. Specifically, the {\tt BMC} describes  Comptonization emission during disk (plane) accretion, while {\tt CompTT} supports coronal (spherical/ plane) Comptonization. 
As a result of this comparison, the better 
model will be used to estimate the BH mass in RXJ1301  (Sect.~\ref{mass_estimate}). In fact, the validity of the {\tt BMC}-based BH mass measurement rests on the assumption that X-ray Comptonization is determined by the {\it Bulk Motion Comptonization} process, not by the processes in the standard hot corona (which is currently the accepted paradigm for X-ray emission of AGNs). 
However, these models are quite closely related, both describes radiative transfer by convolving the thermal spectrum with the Comptonization Green's function, which in both models is a broken power law,

\[G(x,x_0) =
\left\{
\begin{array}{rcl}
(x/x_0)^{-\alpha} , ~~~x\geq x_0,\\
(x/x_0)^{\alpha+3} , ~~x\leq x_0,\\
\end{array}
\right.\]
\noindent where $x = hv/kT_e$, $x_0$ is the breakpoint of $G(x,x_0)$ 
where the two regimes ($x>x_0$ and $x<x_0$) meet each other. Thus, the general model consists of a BBody-like and a Comptonized component (XSPEC models {\tt BMC} and 
 {\tt CompTT}) are the sum of $BB+f\cdot BB*G$. In result,  both models allow for the reproduction of X-ray emission for a wide range of X-ray objects (a more detailed discussion of the models will be presented below in Sect.~\ref{discussion_BMCmodel}).  In this section, we will attempt to assess their possible differences and their suitability for describing the X-ray emission of RXJ1301 in practice.

Namely, we tried to fit the spectra of RXJ1301 using different model formulas: {\tt CompTT}, {\tt BB}, {\tt BMC} and their possible combinations, to choose the most suitable model. 
We also take into account the absorption by neutral matter. 
We investigated the influence of residuals $\Delta\chi$ on the choice of a suitable model for fitting the RXJ1301. 
We also analyzed the best-fit parameters for the spectrum modeling in different models and their acceptability in terms of the most probable physical conditions in RXJ1301.

%
%
\begin{table*}
  \centering 
 \caption{Best-fit parameters of spectral analysis of the {\it XMM-Newton} 
 observations of RXJ1301.  The parameters that are fixed during the fits are denoted with $^{(fixed)}$.} 
 \begin{tabular}
{lllllllr}
      \hline\hline
Model    & Parameter               & $X1^{low}$& $X1^{high}$  &  $X2^{low}$ & $X2^{high}$\\
\hline 
Tbabs    & $N_H$ ($\times 10^{22}$ cm$^{-2}$)     & $4\times 10^{-2~(fixed)}$ & $4\times 10^{-2~(fixed)}$ &$4\times 10^{-2~(fixed)}$ &  $4\times 10^{-2~(fixed)}$ \\
CompTT      & $T_{0}$ (eV)   & 52$\pm$5& 134$\pm$9 &51$\pm$3 &149$\pm$5 \\
             & $\tau$  & 0.93$\pm$0.01 & 1.02$\pm$0.06    &0.86$\pm$0.01 & 1.17 $\pm$0.02   \\
             & N$_{CompTT}^{\dagger}$ &  1.3$\pm$0.2 &  3.0$\pm$0.1 & 2.75$\pm$0.05& 3.97$\pm$0.02\\
Bbody   & $kT_{BB}=T_{0}$ (eV)  & 52$^{(fixed)}$ & 134$^{(fixed)}$ &51$^{(fixed)}$ & 149$^{(fixed)}$\\
            &  $N_{BB}^\dagger$  & 0.01$\pm$0.01& 0.02$\pm$0.02      &0.01$\pm$0.01 & 0.02$\pm$0.01     \\
\hline
           &$\chi^2_{red}$ (dof) &    1.06 (198) &    1.02 (27)& 0.87 (19)&1.26 (29)\\
\hline
\hline
Tbabs    & $N_H$ ($\times 10^{22}$ cm$^{-2}$)     & $4\times 10^{-2~(fixed)}$ & $4\times 10^{-2~(fixed)}$ &$4\times 10^{-2~(fixed)}$ &  $4\times 10^{-2~(fixed)}$\\
BMC       & $T_{s}$ (eV)   &        31$\pm$8&        138$\pm$10 &35$\pm$3 &150$\pm$20 \\
             & $\Gamma_{bmc} = \alpha_{bmc} + 1$& 1.18$\pm$0.01 & 2.71$\pm$0.04 &1.19$\pm$0.03 & 1.59$\pm$0.01\\
             &  $\log(A)$   & 2.3$\pm$0.4 & -3.0$\pm$0.2 &0.5$\pm$0.1& -0.81$\pm$0.04 \\
             & N$_{bmc}^{\dagger}$  & 0.09$\pm$0.01& 1.09$\pm$0.06 & 0.20$\pm$0.01& 0.50$\pm$0.02 \\
\hline
           &$\chi^2_{red}$ (dof) &    0.91 (198)&    1.05 (27)& 0.96 (19)&1.04 (29)\\
\hline
\hline
      \end{tabular}
    \label{tab:fit_table_spectrum}
\begin{flushleft}
$^\dagger$ 
The normalization parameters of {\it BMC} and {\it CompTT} components are in units of $L_{35}/d^2_{10}$ $erg/s/kpc^2$, where $L_{35}$ is the 
source luminosity in units of 10$^{35}$ erg/s; $d_{10}$ is the distance to the source in units of 10 kpc; $\tau$ is the plasma optical depth in the CompTT model. We froze the plasma temperature at $kT_e = 12$ keV because the temperature and optical depth of the Comptonizing plasma are closely related and cannot be simultaneously constrained in the {\it CompTT} model (see details in the text).  
$kT_{0}$ and $kT_s$ are the temperatures of the seed photons. 
$\Gamma_{bmc}$ is the photon index 
of the {\tt BMC}. 
\end{flushleft}
\end{table*}

Let us recall that the parameters of {\tt CompTT} model for a given plasma configuration (flat/spherical) are the temperature of the seed photons   $kT_s$, electron temperature $kT_e$, plasma optical depth $\tau$ (analogous to the spectral index $\alpha$), 
and normalization $N_{CompTT}$. In turn, the parameters of {\tt BMC} model are the spectral index $\alpha$ (analogous to the plasma optical depth $\tau$), the temperature of the seed photons $T_s$, the illuminance factor $log A$ (is related to the Comptonized factor $f={A}/{(1+A)}$), and normalization $N_{BMC}$, defined for both models as the ratio of the source (disk) luminosity $L$ to the square of the distance $d$ (ST09, see Eq.~1 there): 

$$
N_{CompTT, BMC}=\biggl(\frac{L}{10^{39}\mathrm{erg/s}}\biggr)\biggl(\frac{10\,\mathrm{kpc}}{d}\biggr)^2.
$$

In the spectral analysis of XMM-Newton data, we found that the model that includes an accretion disk and a hot coronal component, such as the {\tt Tbabs*(Blackbody + CompTT)} model, provides a good spectral fit to all observations except $X2$ in the 0.2--2.1 keV range, namely, it gives an unacceptable $\chi_{red}^2$ for them. Note that in the $X1^{low}$ state, the signal was dominated by the background at photon energies above 1 keV, so data above these energies were not used; while in the $X1^{high}$ state, the spectra are harder and the source is well detected  up to  at photon energies about 2--3 keV. In the $X1^{low}$ state, spectral data below 0.3 keV were discarded due to remaining calibration uncertainties (see Figs.~\ref{sp_compar_xmm} and \ref{8_spectra_rx}). It should be acknowledged that the {\tt Tbabs*BMC} model provides a better statistical fit to all observations (see Table~\ref{tab:fit_table_spectrum}). To ensure a similar number of degrees of freedom  in the subsequent comparison of both models, we fixed most parameters in the {\tt CompTT} component to typical values \citep[for more details, see][]{Sun13,Giustini24}. 

The best-fit parameters of the combined model {\tt Tbabs*(Blackbody + CompTT)} are $N_H = 4\times 10^{20}$ cm$^{-2}$, $kT_{s} = 52\pm5$ eV, and $\tau = 0.93\pm 0.01$ for $X1^{low}$ ($\chi^2_{red}=1.06$ (198 dof) and $kT_{s} = 134\pm9$ eV, and $\tau = 1.02\pm 0.06$  ($\chi^2_{red}=1.02$ (27 dof) for $X1^{high}$ (see Table~\ref{tab:fit_table_spectrum}). Note that comptonizing plasma are closely related, so we fixed the plasma temperature at 12 keV and obtained constraints on the optical depth. Indeed, assuming the plasma temperature to be a free parameter, we obtain $kT_e = 12~ (\le 20)$ keV and $\tau \le 1$ for the {\it XMM-Newton} 
 spectra. It should be especially noted that both parameters cannot be constrained by the data during the outburst (both parameters contribute equally to the spectrum) and, therefore, cannot be simultaneously constrained in the {\it CompTT} model.  

We then compared the {\tt Tbabs*(CompTT+BB)} model with the {\tt Tbabs*BMC} model.  
The best-fit parameters of the model ({\tt Tbabs*BMC}) are $N_H = 4\times 10^{20}$ cm$^{-2}$, $kT_{s} = 31\pm8$ eV, and $\Gamma = 1.18\pm 0.01$ for $X1^{low}$ ($\chi^2_{red}=0.91$ (198 dof) and $kT_{s} = 138\pm10$ eV, and $\Gamma = 2.71\pm 0.04$  ($\chi^2_{red}=1.05$ (27 dof) for $X1^{high}$ (see Table~\ref{tab:fit_table_spectrum}). 
A comparison of {\tt BMC} with {\tt CompTB+BB} models showed that adding the {\tt BB} component  
does not improve the fits. Its effect on the best-fit results is small  ($N_{BB}\to0.01$ $L_{35}/d^2_{10}$). 
Simulations using {\tt BMC} alone successfully approximate all spectra with acceptable parameters (see Table \ref{tab:fit_table_spectrum}). Although both models are acceptable in terms of $\chi^2$, the {\tt BMC} model is preferable because it has fewer free parameters and demonstrates physically justified values. 
We then used the {\tt Tbabs*BMC} model for the spectral analysis of all data (Table~\ref{tab:fit_table_1301}). 



In all two models, the source flux exceeded 6$\times 10^{-13}$ erg s$^{-1}$ cm$^{-2}$ in the 0.1--2 keV  range  and the corresponding rest frame intrinsic X-ray luminosity is $4.8\times10^{42}$ erg/s. It should be noted that the absorber densities in the columns obtained by fitting the X-ray spectra of RXJ1301 do not exceed the galactic value measured by radio observations in the direction of RXJ1301 \citep{Dickey+Lockman90}, $N_H\sim 9.5\times 10^{19}$cm$^{-2}$. 



For further spectral analysis of RXJ1301 we used the Comptonization model {\tt BMC} modified by neutral absorption {\tt Tbabs} \citep{Wilms00}. 
The parameter of a  {\tt Tbabs} component is a column density $N_H$ (see Table~\ref{tab:parameters_rx}). 
The model reads in XSPEC as {\tt Tbabs*BMC}.

%
%

\begin{figure}
\includegraphics[scale=0.42,angle=0]{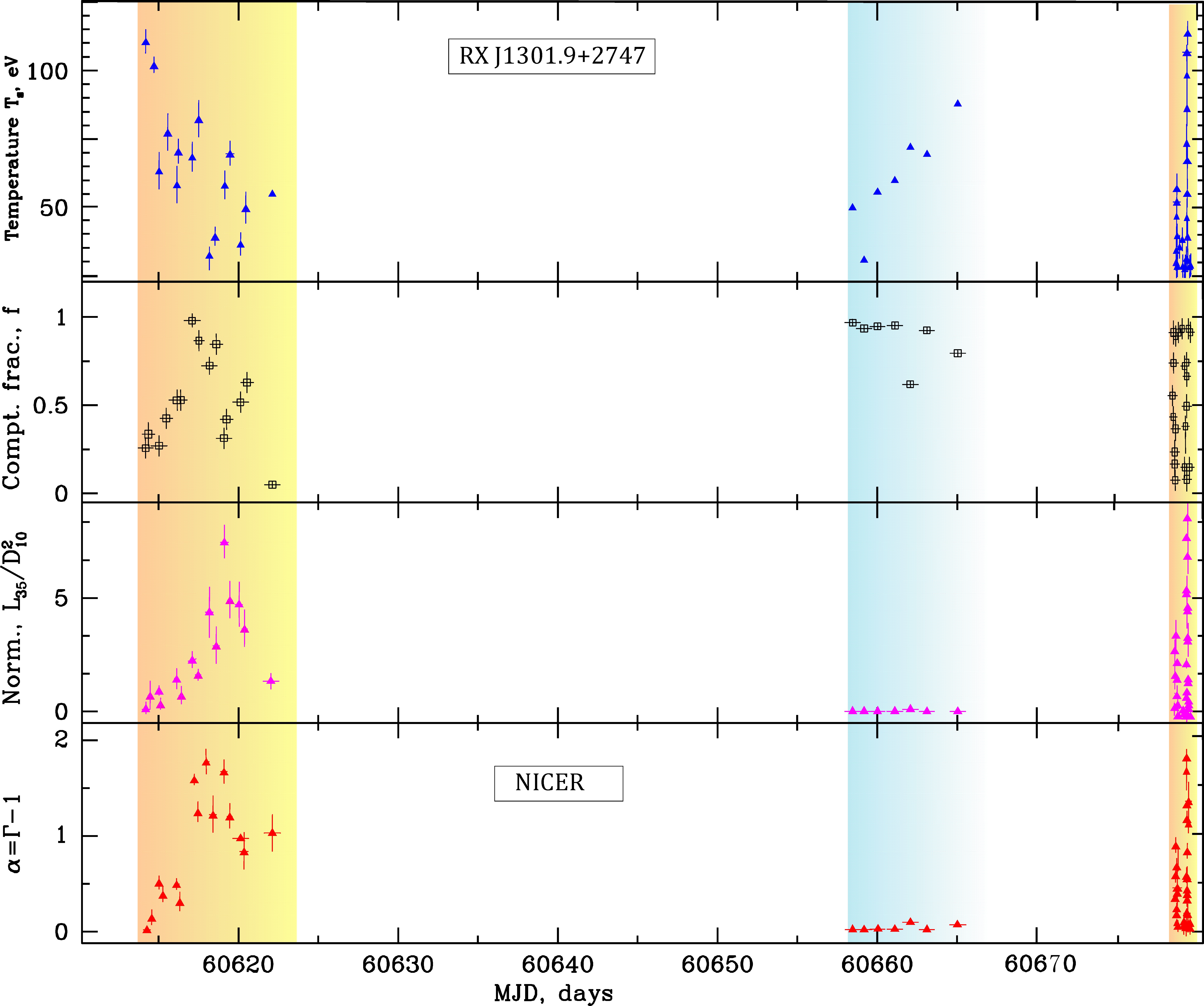}
\caption{
From top to bottom: 
seed photon temperature $kT_s$ (in eV), Comptonized fraction $f$, and {\tt BMC} normalization $N_{bmc}$ during 2024--2025 outburst events of RXJ1301 using NICER observations. In the bottom panel, we present an evolution of the spectral index $\alpha = \Gamma - 1$. The interval of increased outburst activity is marked by an orange vertical stripe, although the phase of low-amplitude flares during the interval of decreased activity of RXJ1301 is indicated by blue vertical stripe. 
}
\label{fraq_rx}
\end{figure}

\begin{figure}
\includegraphics[width=8.5cm]{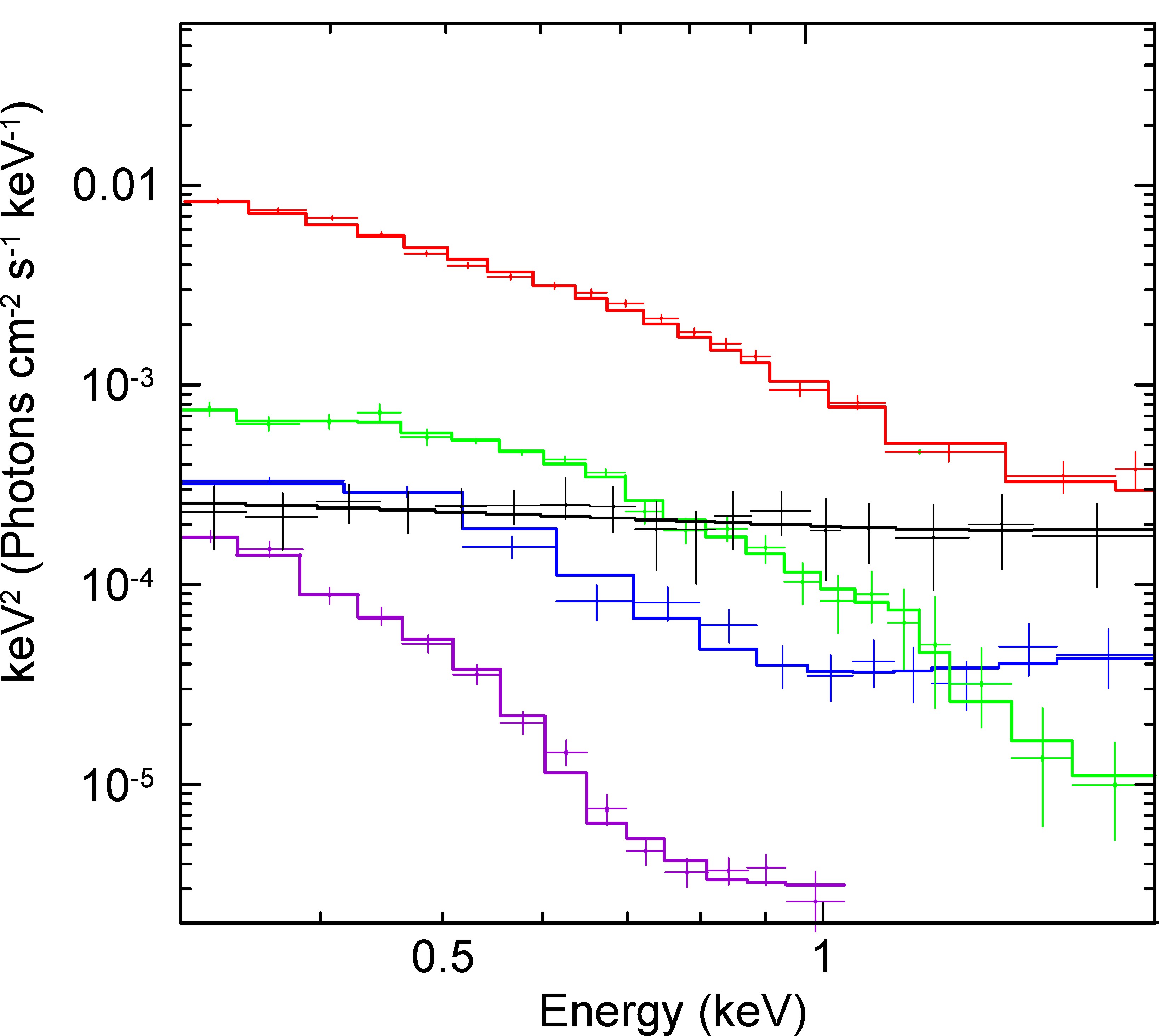}
\caption{Five 
representative $E*F(E)$ spectral diagrams that are related to different spectral states of RX~J1301.9+2747 using the model {\tt Tbabs*BMC}. 
The data are taken using 
{\it NICER} observations $Nc3^{high}$ (red), $Nc3^{low}$ (black), by {\it XMM-Newton} $X1^{high}$ (green), $X1^{low}$ (violet), and by the {\it ROSAT} observation $R2$ (blue). 
The data are denoted by  crosses, while the spectral model is shown by histograms for each spectral state.
}
\label{8_spectra_rx}
\end{figure}

Our spectral analysis of the RXJ1301 observations by {\it ASCA, NICER, XMM-Newton} and {\it ROSAT} 
indicates  that  the source  spectra can be  well reproduced by the  above mentioned model. 
In Figure \ref{fraq_rx} we show how the parameters $T_s$, $f$, $N_{bmc}$, and $\Gamma$ change during the { \it NICER} observations. 
From this Figure it is clear that during the interval of increased luminosity all parameters show noticeable variability (orange vertical stripe), although during the period of relatively low luminosity some of  these parameters are almost constant.
In particular, 
orange vertical strip indicates 
high-amplitude outburst variability according to the normalization, (0.3--6.5) $L_{35}/D_{10}^2$, 
and the 
blue vertical strips mark 
low-amplitude variability at (0.03--0.3) $L_{35}/D_{10}^2$ 
according to {\it NICER} data. 

\begin{figure}[ptbptbptb]
\centering
\includegraphics[scale=0.45,angle=0]{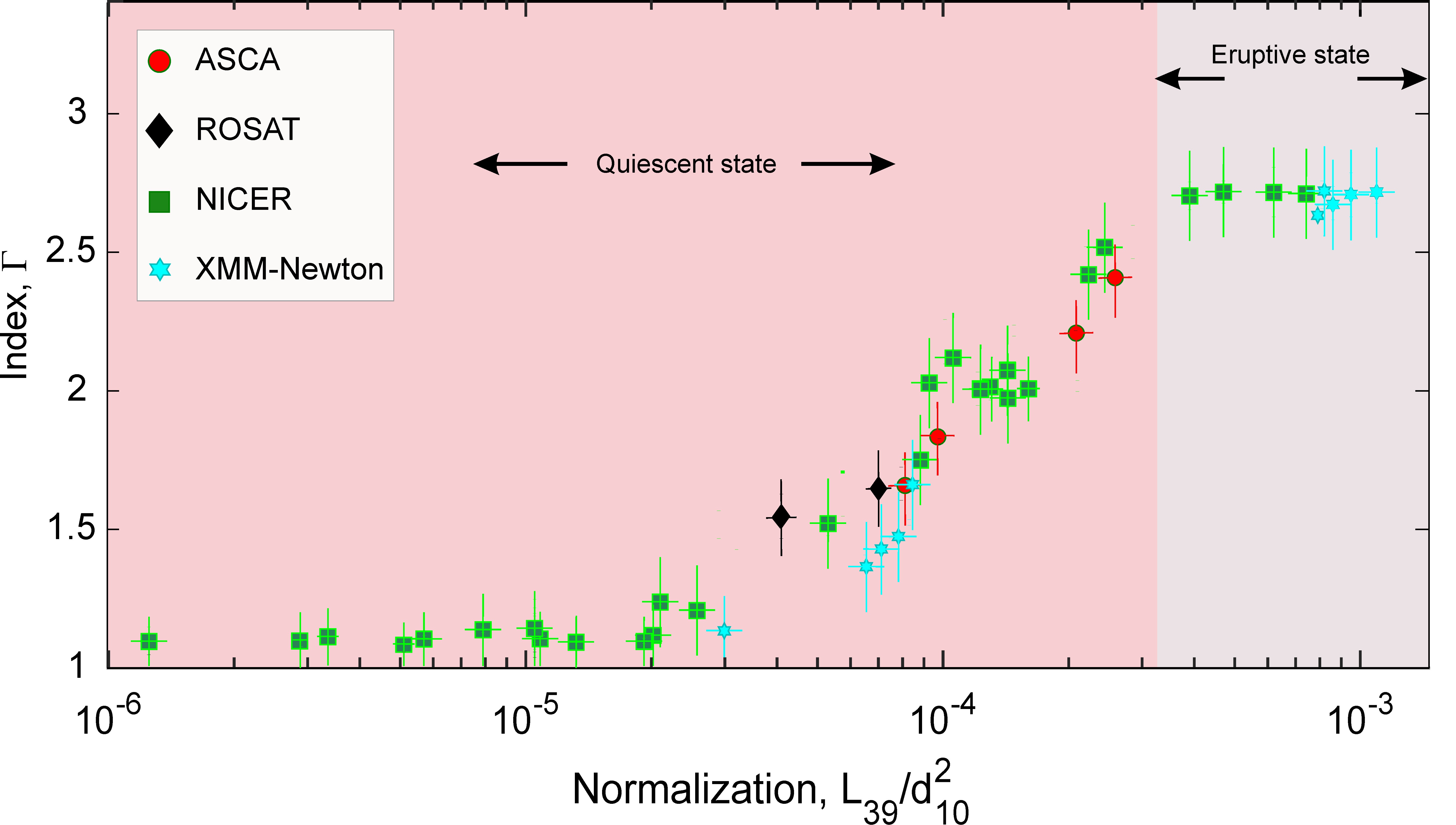}
\caption{Photon index, $\Gamma$ plotted versus BMC normalization (which is proportional to $\dot M$) for RXJ1301.
}
\label{saturation_all}
\end{figure}

Our results using flux-resolved methodology indicate three $\Gamma - N$ regimes in RX1301: i) low normalization for low and 
medium normalization for medium count rate observations (QS states); and ii) high normalization for high count rate events (ES states). Furthermore, the $\Gamma-N$ diagram shows a smooth trajectory (Fig.~\ref{saturation_all}). In this case, we grouped the {\it NICER, XMM-Newton} and {\it ASCA} spectra into two characteristic bands according to the count rate for each observation (see Section \ref{spectral analysis}) and fitted the combined spectra of each band using the XSPEC package. 

Overall evolution of the spectral shape for RXJ1301 during the transition from the quiescent to the outburst state is presented in Fig.~\ref{8_spectra_rx}. 
Here we present observations in the QS  ($Nc3^{low}, X1^{low}$), where the spectrum of RXJ1301 is dominated by the hard emission component, as well as by soft humps ($Nc3^{high}, X1^{high}$). 
 In this figure, we have show representative spectra that have a high signal-to-noise ratio. 
 All of them are well reproduced by the Comptonization model. For example, the best-fit parameters for the spectrum of  $Nc3^{low}$ are $\Gamma$=1.00$\pm$0.06 and $T_s$=58$\pm$9 eV 
(reduced $\chi^2$=0.89 for 31 dof). 
While the best-fit parameters for the spectrum of $ Nc3^{high}$ are $\Gamma$=2.21$\pm$0.02 and $T_s$=60$\pm$9 eV (reduced $\chi^2$=0.91 for 31 dof). Details for other observations see in Table~\ref{tab:fit_table_1301}. 

We clearly see that the spectral state change 
 in RXJ1301 is accompanied by a change of the seed photon temperature $kT_s$ between 30 and 120~eV and an increase in the $\Gamma$ index from 1 to 2.7 (Figs.~\ref{fraq_rx}, \ref{saturation_all} and \ref{scaling_rx}). We have detected complex X-ray variability (Fig.~\ref{fraq_rx}) and established detailed patterns of X-ray spectral transitions in RXJ1301 based on flux-resolved {\it NICER} observations during 2024. These patterns are consistent with previous observations of this source \citep[e.g.,][]{Song20}, where the X-ray light curve of the source also shows diverse behavior over a wide range of time scales: from hours to years \citep[e.g.,][]{Song20,Zhao22}. 
Here we discuss the source variability on hourly time scales and demonstrate an evolution of the model parameters for all analyzed spectral transitions in the flares in Fig. \ref{fraq_rx}. Thus, the interval of enhanced flaring activity (MJD 60614--60622), marked by the orange vertical stripe in Fig.~\ref{fraq_rx}) is characterized by variability of $kT_s$ (30 -- 112 eV), the fraction  of the Comptonized photons  $f$ (0 -- 1), the BMC normalization, $(0.03-7.7)\times L_{35}/D_{10}^2$, and the photon index $\Gamma$ (1 -- 2.7). In contrast, the interval of the quiescent state of the source (MJD 60658--60666, marked by the blue vertical stripes) is characterized by weak variability of $kT_s$ (30 -- 88 eV), the fraction  of the Comptonized photons  $f$ (0.6 -- 1), the BMC normalization, $(0.05 - 0.2)\times L_{35}/D_{10}^2$, and the photon index $\Gamma$ (1 -- 1.1). Spectral analysis results are reported in Table~\ref{tab:fit_table_1301}.

It is worth noting that the {\it NICER} observations of RXJ1301 cover the phase of increasing soft excess in the source spectrum at 0.2--1 keV against the background of hard emission (1.5--2 keV) in the QS (see Figs.~\ref{2_spectra} and \ref{8_spectra_rx}). The outburst state of RXJ1301, when it passed to a softer state, is also well covered by the {\it NICER, ASCA} and {\it ROSAT} observations. It is obvious that in this state the spectrum of the source is dominated by soft radiation (for example, the violet spectrum $X1_{low}$ and  the blue spectrum $R2$ in Fig.~\ref{8_spectra_rx}). 


%
%

In Figure~\ref{saturation_all} we demonstrate the dependence of the photon index $\Gamma$ on the BMC normalization $N$ for {\it ASCA} (red), {\it ROSAT} (black), {\it XMM-Newton} (cyan) and {\it NICER} (green) observations. It is obvious that the $\Gamma-N$ relation  shows a monotonic increase of $\Gamma$ with a normalization increase  followed by  a constancy (the so-called saturation) of $\Gamma$ at large values of $N$. Such a behavior of $\Gamma$ is expected for objects containing BHs~\citep{tz98}, and has already been observed in many sources containing SMBHs \citep{TS17} and BHs of both stellar \citep{ST10,STS14} and intermediate masses \citep{TS16}. This effect, now also detected in RXJ1301, was used by us to estimate a BH mass in this object.

It is interesting to note that the $\Gamma-N$ track captures the region of monotonic growth of $\Gamma-N$ for small $N$ (which is proportional to $\dot M$ and the initial region of index saturation in the $\Gamma-N$ diagram for higher $N$ (see Fig.~\ref{saturation_all}). This is consistent with the noticeable variability in the 0.3--2 keV flux of RXJ1301 during the quiescent state (only outside of QPE-outbursts, see Fig.~9 in Giustini et al. (2024) based on {\it EXOSAT, ROSAT, XMM-Newton}, and {\it Chandra} data). This provides a basis for applying the scaling method even when using such QS states (pink region in Fig.~\ref{saturation_all}).

%
%
\begin{table*}
  \centering 
 \caption{Best-fit parameters for the spectral analysis of  {\it ROSAT}, 
 {\it NICER}, 
 {\it ASCA}, 
and {\it  XMM-Newton} 
 RXJ1301 observations $^{\dagger}$.
}
\resizebox{1.0\textwidth}{!}{%
 \begin{tabular}{llllll|lllllllccccccccccc}
      \hline\hline
ID& $\alpha=$&$T_s$,& $\log(A)$ & N$_{com}^{\dagger\dagger}$ & 
$\chi^2_{red}$ &ID& $\alpha=$&$T_s$,& $\log(A)$ & N$_{com}^{\dagger\dagger}$ &$\chi^2_{red}$ \\
           & $\Gamma-1$ &eV& &               &
  (dof)&           & $\Gamma-1$ &eV& &               &   (dof)\\
\hline
\hline
R1 &  0.54$\pm$0.08  & 67$\pm$1   &    -0.16$\pm$0.05 &40.97$\pm$0.06 & 0.85 (39) & Nc16$^{high}$&    1.69$\pm$0.05 & 149 $\pm$8     &  0.14$\pm$0.05 &390$\pm$10  & 0.99 (39)\\
R2&  0.64$\pm$0.01  & 72$\pm$5   &    -0.89$\pm$0.03 &70.04$\pm$0.09 & 1.17 (35) &Nc17$^{low}$&    0.15$\pm$0.09 & 65 $\pm$3     &  0.19$\pm$0.04 &10.5$\pm$0.8  & 1.03 (34)\\
Nc1$^{low}$&   0.08$\pm$0.01 &  110$\pm$8&  0.35$\pm$0.09 & 13.2$\pm$0.09 &0.76 (34) &  Nc17$^{high}$&    1.71$\pm$0.04 & 150$\pm$10     &  0.21$\pm$0.01 &570$\pm$8  & 1.08 (39) \\
Nc1$^{high}$&   0.09$\pm$0.01 &  112$\pm$10&  0.27$\pm$0.07 & 19.2$\pm$0.1 &0.89 (34)& Nc18$^{low}$&    0.12$\pm$0.01 & 59$\pm$6     &  0.23$\pm$0.05 &10.5$\pm$0.8  & 0.92 (34)\\
Nc2$^{low}$&    0.09$\pm$0.05&   61$\pm$9 &  0.15$\pm$0.06 &1.25$\pm$0.09& 1.10 (37)&  Nc18$^{high}$&    0.71$\pm$0.09 & 148$\pm$5     &  0.18$\pm$0.02 &620$\pm$10  & 0.96 (39)\\
Nc2$^{high}$&    1.01$\pm$0.05&   63$\pm$6 &  0.11$\pm$0.05 &130$\pm$8&1.12 (37)& Nc19$^{low}$&    0.11$\pm$0.05 & 57$\pm$6     &  0.16$\pm$0.03 &10.5$\pm$0.8  & 1.04 (34)\\
Nc3$^{low}$&  1.00$\pm$0.06 &58$\pm$9 & -0.32$\pm$0.07 &121$\pm$20 & 0.89 (31)&Nc19$^{high}$&    0.46$\pm$0.05 & 110$\pm$5     &  0.22$\pm$0.04 &220$\pm$10  & 0.89 (34)   \\
Nc3$^{high}$&  1.21$\pm$0.02 &60$\pm$9 & -0.54$\pm$0.01 &148$\pm$23 & 0.91 (31) & Nc20$^{low}$&    0.08$\pm$0.01 & 58 $\pm$3     &  0.11$\pm$0.04 &0.08$\pm$0.02  & 0.96 (34) \\
Nc4$^{low}$&   0.24$\pm$0.02 & 65$\pm$7  &-0.23$\pm$0.07 &21$\pm$2 & 1.08 (37)& Nc20$^{high}$&    0.62$\pm$0.08 & 91$\pm$8     &  0.15$\pm$0.01 &57.5$\pm$0.8  & 0.98 (35)\\
Nc4$^{high}$&   0.95$\pm$0.02 & 69$\pm$4  &-0.41$\pm$0.02 &143$\pm$14 & 1.05 (37)& Nc21&    0.12$\pm$0.09 & 56 $\pm$3     &  0.2$\pm$0.1 &10.5$\pm$0.8  & 1.02 (39)  \\
Nc5$^{low}$&   0.51$\pm$0.05 & 30$\pm$8  & 0.11$\pm$0.09 &29$\pm$8 & 1.08 (39)& Nc22&    0.13$\pm$0.08 & 34 $\pm$2     &  0.28$\pm$0.09 &10.5$\pm$0.8  & 1.09 (39) \\
Nc5$^{high}$&   1.73$\pm$0.06 & 31$\pm$9  & 0.13$\pm$0.04 &385$\pm$70 & 1.14 (39)& A1&   1.41$\pm$0.01 & 30$\pm$5  &  -2.3$\pm$0.1  &244$\pm$50  & 1.11 (160)\\
Nc6$^{low}$&   0.50$\pm$0.03 & 65$\pm$8      &-0.96$\pm$0.08 &53$\pm$3 &  0.87 (35)& A2&    1.20$\pm$0.01&  39$\pm$4 &   -2.23$\pm$0.04 &208$\pm$35 &0.78 (119)\\
Nc6$^{high}$&   1.70$\pm$0.01 & 69$\pm$7      &-0.88$\pm$0.03 &742$\pm$20 &   0.84 (35)& A3&   0.73$\pm$0.06 & 29$\pm$6  &  -2.14$\pm$0.06 &94$\pm$9  &1.13 (126)\\
Nc7$^{low}$&   0.41$\pm$0.01 &31$\pm$7   &-1.08$\pm$0.07 &32$\pm$3 & 0.91 (34)&  A4&   0.81$\pm$0.01 & 32$\pm$9  &  -2.28$\pm$0.04 &98$\pm$8     &1.09 (126)\\
Nc7$^{high}$&   1.01$\pm$0.02 &33$\pm$5   &-1.15$\pm$0.06 &210$\pm$30 & 0.89 (34)& X1$^{low}$&0.51$\pm$0.01 & 31$\pm$8 &   2.3$\pm$0.4& 83$\pm$10  &  0.91 (198)\\
Nc8&   1.17$\pm$0.08 & 57$\pm$1     & -0.94$\pm$0.05 &101$\pm$6  &0.85 (39)& X1$^{high}$&1.71$\pm$0.04   & 138$\pm$10 &   -3.0$\pm$0.2& 1094$\pm$60 & 1.05 (27) \\
Nc9&   0.08$\pm$0.01 & 50$\pm$18     &  2.00$^{\dagger\dagger\dagger}$ &  2.9$\pm$0.6&  0.99 (33) & X2$^{low}$&0.43$\pm$0.03 & 35$\pm$3 &   0.5$\pm$0.1& 74$\pm$10  &  0.96 (19) \\
Nc10&    0.07$\pm$0.02  &30$\pm$2      & 2.00 $^{\dagger\dagger\dagger}$ &5.1$\pm$0.2  & 0.97 (30) & X2$^{high}$&1.65$\pm$0.01   & 150$\pm$20 &   -0.81$\pm$0.04& 790$\pm$20 & 1.04 (29)\\
Nc11&    0.10$\pm$0.01  &58$\pm$5      &-0.18$\pm$0.02 &11$\pm$2  & 1.02 (33)& X3$^{low}$&0.66$\pm$0.01 &38$\pm$7 & -0.2$\pm$0.1& 85$\pm$3 &  1.08 (13)\\
Nc12&    0.10$\pm$0.01 & 60$\pm$1     & -0.08$\pm$0.01 &20$\pm$1  & 0.85 (34)& X3$^{high}$&1.70$\pm$0.06   & 110$\pm$4 &   -2.1$\pm$0.2& 860$\pm$60 & 1.06 (93) \\
Nc13&    0.13$\pm$0.01 & 71$\pm$8     &  0.25$\pm$0.02 &7.9$\pm$0.6  & 0.93 (34)&X4$^{low}$&0.39$\pm$0.01 & 25$\pm$8 &   -1.4$\pm$0.4& 65$\pm$6  &  0.95 (13) \\
Nc14&    0.10$\pm$0.01 & 69$\pm$6     &  0.14$\pm$0.03 &5.7$\pm$0.9  & 0.94 (33)& X4$^{high}$&1.70$\pm$0.05   & 100$\pm$2 &   -2.6$\pm$0.2& 810$\pm$60 & 1.08 (93)\\
Nc15&    0.12$\pm$0.07 & 88 $\pm$4     &  0.22$\pm$0.04 &10.5$\pm$0.8  & 0.89 (34)& X5$^{low}$&0.09$\pm$0.01 & 25$\pm$8 &   -1.9$\pm$0.4& 30$\pm$6  &  0.93 (13)\\  
Nc16$^{low}$&    0.13$\pm$0.05 & 58 $\pm$3     &  0.23$\pm$0.06 &10.5$\pm$0.8  & 0.95 (34)& X5$^{high}$&1.71$\pm$0.08   & 95$\pm$2 &   -2.3$\pm$0.2& 800$\pm$60 & 1.01 (93) \\
\hline
 \end{tabular}
}
    \label{tab:fit_table_1301}
\begin{flushleft}
$^\dagger$ The spectral model is  {\tt Tbabs* BMC}. 
$^{\dagger\dagger}$ Normalization parameters of {\it {\tt BMC}} component is in units of $L_{33}^{\rm soft}/d^2_{10}$, $erg/s/kpc^2$, 
where $L_{33}^{\rm soft}$ is  the soft photon  luminosity in units of 10$^{33}$;  
$d_{10}$ is the distance to the source in units of 10 kpc. 
$^{\dagger\dagger\dagger}$ When parameter log(A)$>>$1, it is fixed to a value 2.0. 
$N_H$ varied slightly around the value  
4$\times 10^{20}$ cm$^{-3}$ (see also \cite{Sun13}). 
\end{flushleft}
\end{table*}


\subsection{A BH mass estimate in RX~J1301.9+2747}
\label{mass_estimate}

We used the 
scaling method (ST09), 
which is widely used to estimate a BH mass, $M_{BH}$ surrounded by an accretion disk~\citep{TS24,TSC23,STU18,SCT18,STV17}. It is assumed that under certain conditions the X-ray emission characteristics of the inner region of the disk are similar for different BHs of different masses \citep{SS73}, which is consistent with empirical data \citep[see, e.g., ][]{TS23,TSM25}. Such a situation is realized, for example, for a BH involved in a binary system, and the mass of such a BH affects the size of the accretion disk surrounding it and its luminosity \citep{SS73}:
\begin{equation}
L =G\frac{ \dot M M_{BH}}{R_{\ast}} ,
\label{mass}
\end{equation}
\noindent where $G$ is the gravitational constant, $\dot M$ is the mass accretion rate, 
$R_{\ast} = r_{\ast} \times R_S$ is the effective radius at which the main energy release in the disk occurs, and $R_S = 2GM_{BH}/c^2$ is the Schwarzschild radius. Under this assumption, the mass for one BH object, $M_{BH}$  (``{\it target}'' object, $M_{t}$) can be scaled relative to the mass of another BH object, `$M_{r}$ (`{\it reference}'' object):  
\begin{equation}
M_t=\frac{M_r}{s_N}\left(
\frac{d_t}{d_r}
\right)^2 f_G,
\label{mass_target1}
\end{equation}
where $d_r$ and $d_t$ are the distances to the ``{\it reference}'' and ``{\it target}'' sources, $f_G = \cos(i_r)/\cos (i_t)$ is a geometric factor that takes into account the different inclinations of the disk to the observer  for the ``{\it reference}'' ($i_r$) and ``{\it target}'' ($i_t$) objects, and $s_N = N_r/N_t$ is a scale factor that takes into account the different X-ray fluxes from the ``{\it reference}'' $N_r$ and ``{\it target}'' $N_t$ objects. 
When calculating $s_N$, the fact is used that the X-ray spectrum of all BH sources changes in a similar way during their flares, namely: the photon index $\Gamma$ increases with the X-ray flux $N$, reaching the ``saturation phase'' of the index $\Gamma$ at the maximum of the flare \citep{st09,TS23}. Then the fluxes $N_r$ and $N_t$ for the ``{\it reference}'' and ``{\it target}'' objects are measured on the $\Gamma-N$ diagram at the corresponding points at which the $\Gamma-N_r$ and $\Gamma-N_t$ correlations reach the saturation level in the $\Gamma$. To best satisfy the similarity conditions of scalable objects, it is necessary to find an ``{\it reference}'' object for the ``{\it target}'' BH that is an accreting BH and has the same saturation level $\Gamma_{sat}$ as the ``{\it target}'' BH on the $\Gamma-N$ correlation diagram. It should be especially noted that the activity of the object (X-ray flares) favors the applicability of this method.

%
%

To determine the distance to RX~J1301.9+2747 we used Ned Wright's Javascript Cosmology Calculator\footnote{https://www.astro.ucla.edu/~wright/CosmoCalc.html} in frame of the LambdaCDM cosmological model \citep{Wright06}, 
 $d_{1301}\sim$ 102.6 Mpc, 
for the redshift $z_{1301} = 0.02358$ \citep{Giustini20}. 
%
%

\begin{figure*}
\centering
  \includegraphics[width=17cm]{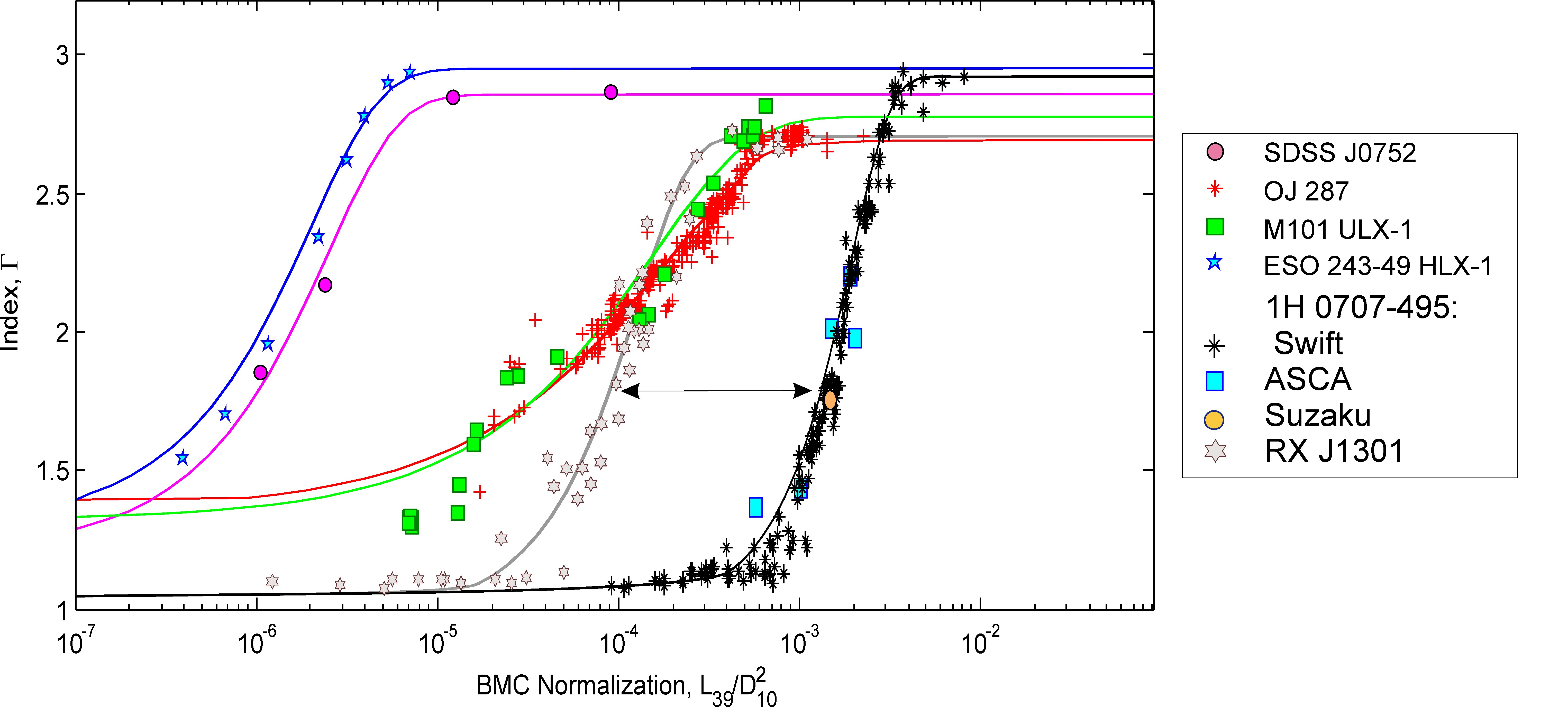}
\caption{Scaling of photon index $\Gamma$ for RXJ1301 (with grey line for target source) with HLX--1~ESO 243--49, 
OJ 287, 
and 1H~0707--495 (as reference sources). 
 The horizontal black arrow highlights the fact that the correlations of the target and reference sources are similar. The only difference is the BMC normalization due to the different $M_{BH}/d^2$ ratios.
}
\label{scaling_rx}
\end{figure*}


For the appropriate scaling, we used HLX--1~ESO 243--49, 
OJ 287, 
and 1H~0707--495 
as reference sources, which also show the index saturation effect at the same $\Gamma$ level (Fig.~\ref{scaling_rx}) as in RXJ1301 (the target source). Meanwhile, the BH mass, inclination and distance are well known for the reference sources (Table~\ref{tab:par_scal_rx}).

In Figure \ref{scaling_rx} we demonstrate how the photon index $\Gamma$ evolves with normalization $N$ (proportional to the mass accretion rate $\dot M$) in RXJ1301 (grey line, {\it target} source) 
and HLX--1~ESO 243--49 (blue line), 
OJ~287 (red line), 
and 1H~0707--495 (black line, 
 {\it reference} sources),  where  $N$  is presented in units of $L_{39}/d^2_{10}$ ($L_{39}$ is the source luminosity in units of $10^{39}$ erg/s and $d_{10}$ is the distance to the source in units of 10 kpc). We can see from this Figure 
that these sources have almost the same index saturation level $\Gamma$. 
We estimated a BH mass for RXJ1301 using the scaling approach  (see e.g.,  ST09). In Figure~\ref{scaling_rx} we illustrate  how the scaling method works shifting one correlation versus another.     From these correlations we could estimate $N_t$, $N_r$ for RXJ1301 and   for the reference sources (see Table~\ref{tab:par_scal_rx}). 
We define the value $N_t=4.04\times10^{-4}$ ($N_t$ is measured in units of $L_{39}/d^2_{10}$) at the beginning of the $\Gamma$-saturation part of $\Gamma-N$ correlation (see Fig.~\ref{tab:par_scal_rx}, ST09, \cite{STS14,TS16,tsei16b,ts09})

\begin{figure}
 \centering
\includegraphics[width=8.5cm]{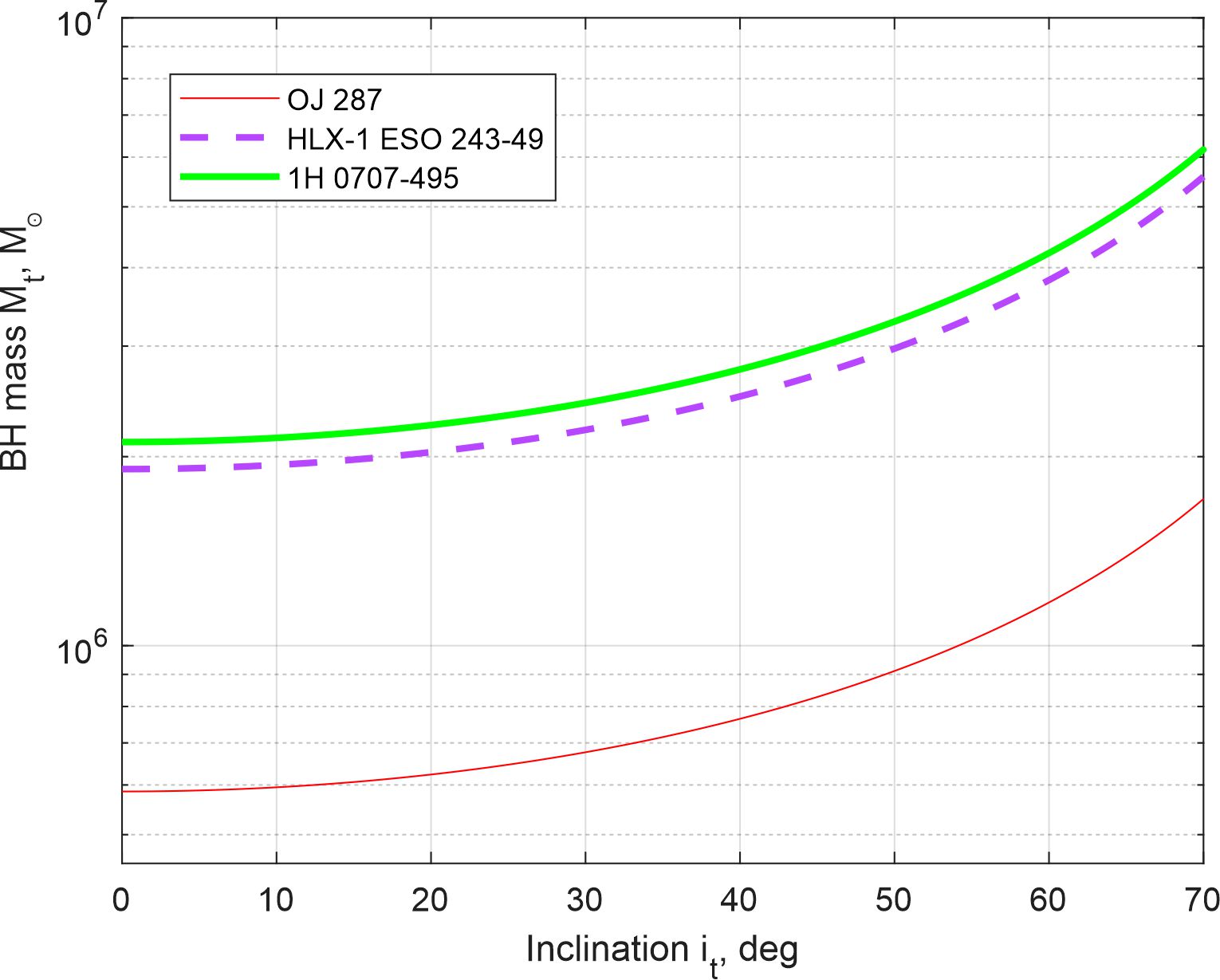}
\caption{A BH mass $M_t$ in RXJ1301 versus the inclination angle $i_t$ (from 0 to 70 deg) using 
OJ~287 (red), 
HLX--1~ESO 243--49 (violet), 1H~0707--495 (green) 
as reference sources. 
It is clear that the BH mass in RXJ1301 is, on average, more than $10^6$ M$_{\odot}$ for inclinations $i > 50^{\circ}$, and less than $10^6$ M$_{\odot}$ for inclinations $i < 50^{\circ}$ 
 (using OJ~287 as the reference) and less than $3\times 10^6$ M$_{\odot}$ for inclinations $i < 50^{\circ}$ (using HLX--1~ESO 243--49 and 1H~0707--495 as the references). 
}
\label{m_i_RX}
\end{figure}

A value of   $f_G$ 
for the {\it target} and {\it reference} sources  can be obtained  using a trial inclination  range for RX~J1301 $i_t=0^{\circ} - 70^{\circ}$  
(see Table~\ref{tab:par_scal_rx} and  
 Fig.~\ref{m_i_RX}). 
In choosing this 
inclination range 
 we are guided by a physically motivated constraint, based on current data and the unified AGN model. Specifically, we believe that the inclination $i_t$ cannot be high, given the little absorption in the X-ray source spectrum; otherwise, the disk would be obscured by the dusty torus.


In result, 
we estimate $m_t$ (see Table~\ref{tab:par_scal_rx}) 
 in RXJ1301 for $i_t=0^{\circ} - 70^{\circ}$ using Eq.~(\ref{mass_target1}). 
Taking HLX--1~ESO 243--49, 
OJ 287, 
and 1H~0707--495 
as reference sources (Fig.~\ref{scaling_rx}), we find that the BH mass in RXJ1301 is 
 from $5.93\times(1\pm 0.26)\times10^5$ M$_{\odot}$ to $6.01\times(1\pm 0.26)\times10^6$ M$_{\odot}$. 
 We present all these results in Table~\ref{tab:par_scal_rx}. To obtain this estimate  with appropriate error bars, we need to consider  error bars for $m_r$ and $d_r$ assuming, in the first approximation, errors for  $m_r$ and $d_r$ only.
 We rewrote Eq. (\ref{mass_target1}) as
 \begin{equation}
m_t(1+\Delta m_t/m_t)= f_G\frac{m_r}{s_N} \frac{d_t^2}{d_r^2}(1+\Delta m_r/m_r)(1+ 2 \Delta d_r/ d_r).
\label{mass_target_expand}
\end{equation}

Thus, we obtained errors for the $m_t$ determination (see Table \ref{tab:par_scal_rx}, second column for the {\it target} source), such that 
\begin{equation}
\Delta m_t/m_t \sim \Delta m_r/m_r + 2 \Delta d_r/ d_r.
\label{mass_target_errors}
\end{equation} 
We present all these results in Table~\ref{tab:par_scal_rx}.

In order to calculate  the  dispersion ${\mathcal D}$ of the arithmetic  mean  $\bar{ m_t}$ for a BH  mass estimate using different reference sources ${\mathcal D}$  (see   Table \ref{tab:par_scal_rx}), one should keep in mind  that 
\begin{equation}
{\mathcal D} (\bar{m_t})= D/n,
\label{dispersion_mean}
\end{equation} 
where $D$ is the dispersion of $m_r$  using  each of the reference sources and $n=3$ is a number of  the reference sources. As a result we determined that  the mean deviation  of the arithmetic mean  
\begin{equation}
\sigma  (\bar{m_t})= \sigma/\sqrt{n}\sim 0.26 
\label{sigma_arithm_mean}
\end{equation}
 and finally  we came to the following conclusion (see also Table \ref{tab:par_scal_rx}):
 \begin{equation}
\bar{m_t} \sim5.93\times(1\pm 0.26)\times10^5 \div 6.01\times(1\pm 0.26)\times10^6 
\label{arithm_mean}
\end{equation}
in the solar masses $\rm M_{\odot}$ for $i_t$ from 0$^{\circ}$ to 70$^{\circ}$.
In fact, determination of  a BH mass in RXJ1301 depends on information of  the inclination $i_t$, so we plotted a BH mass versus inclination $i_t$ in Fig.~\ref{m_i_RX} for source inclinations from 0 to 70 degrees. According to this plot, a BH mass in RXJ1301 
roughly ranges from $5\times 10^5$ to $6\times 10^6$ solar masses.

\begin{table*}
 \caption{BH mass scaling for RX~J1301.9+2747.}
 \label{tab:par_scal_rx}
 \centering 
\resizebox{1.\textwidth}{!}{%
 \begin{tabular}{lllrrll}
 \hline\hline                        
\rowcolor[gray]{.9}Reference sources   & $m_r$,  M$_{\odot}$       & $i_r^{(a)}$, deg  & $N_r$,               $L_{39}/d^2_{10}$      & $d_r^{(b)}$, Mpc  \\

      \hline
HLX--1~ESO 243--49$^{(1)}$ & $(7.2\pm 0.7)\times 10^4$ & 75 & $4.2\times 10^{-6}$ & $ 95\pm 10$\\
OJ 287$^{(3)}$ & $(1.25 \pm 0.5)\times 10^8$ & $50$ & $6.2\times 10^{-4}$ & $1037\pm 10$\\
1H~0707--495$^{(6)}$  & $(7 \pm 2)\times10^7$ & $55$ & $3.5 \times 10^{-3}$ & $\sim$160 \\
 \hline\hline                        
\rowcolor[gray]{.9}Target source   & $m_{t}/cos(i_t)$,  M$_{\odot}$  & $i_t^{(c)}$, deg  & $d_t^{(b)}$, Mpc&  Comments  \\
      \hline
RX~J1301.9  & $\sim5.93\times(1\pm 0.45)\times10^5$ &  0 -- 70 &102.6    &    that  using  HLX--1~ESO 243--49  as a ref. source\\
RX~J1301.9  & $\sim1.92\times(1\pm 0.45)\times10^6$ &  0 -- 70 &102.6    &    that  using  OJ 287  as a ref. source\\
RX~J1301.9 & $\sim2.01\times(1\pm 0.45)\times10^6$ &  0 -- 70 &  102.6    &    that  using  1H~0707--495 as a ref. source\\ 
RX~J1301.9  & Final  estimate                                      & & 102.6   &  \\
           &  $m_t \sim4.3\times 10^5 - 7.5 \times 10^6$ M$_{\odot}$  &    &      & for $i_t$ from 0 to 70 deg$^{(c)}$  \\
\hline
 \end{tabular}
}
\begin{flushleft}
(a)  System inclination in the literature; 
(b) source distance found in the literature,  and  
 (c) source inclination range, which is adopted as a plausible range based on the AGN unified model assumption. 
\\
(1) \cite{tsei16b}; 
(2)  \cite{TS16};  
(3) \cite{TSC23}; 
(5)  \cite{tsei16b}; and  
(6) \cite{TSM25} . 
\end{flushleft}
 \end{table*}

\section{Discussion \label{discussion}}

\subsection{On the X-ray variability and QPE activity of RXJ1301 (1991--2025) \label{discussion_QPE-activity}}

To understand the QPE phenomenon in AGNs and, in particular, in RXJ1301, it is very important to know the time of the first QPE event, its duration, amplitude, and pattern in a given AGN, as well as the transitions between activity states of this AGN. The historical evolution of X-ray emission from RX J1301.9+2747, taking into account all the X-ray data used, allows us to trace the occurrence, intensity, and character of QPE (or QPE-like) events over the past 36 years. We can state that RXJ1301 did not exhibit QPEs in 1991~\citep[see Fig.~4 in][and Fig.~\ref{ev_rx_xmm} here]{Dewangan00}, 1993, and 1994 (Fig.~\ref{ev_rx_xmm}). X-ray activity with QPE events was noticed in RXJ1301 only starting from 2000 \citep{Sun13,Song20,Giustini24}. QPE events continued in RXJ1301 in 2009,  
2019, 2020, 2022 \cite{Giustini24}, 2024, and 2025 (Fig.~\ref{ev_rx_xmm}). These events either appeared or intensified from 
2000 to the present. These events appeared to be stable from 2019 to 2022 and exhibited periodic variability \citep{Sun13,Giustini24}. In the last two years (2024--2025), the object's QPE variability has been rather episodic (see bottom panels in Fig.~\ref{ev_rx_xmm}). 
Overall, X-ray observations of RXJ1301 from 2000 to the present  time  provided  the strong evidence that QPEs were present in this source for the last 26 years.
 In light of such QPE activity, the hypothesis that RXJ1301's activity could be caused by tidal disruptions of various stellar-mass objects by black holes \citep{Bandopadhyay26} is not inconsistent with observations.
In this case, periodically recurring
QPEs may be caused by the incomplete disruption of a massive stellar object (observed in 2019--2022), although 
episodic QPEs may 
be due to the complete disruption of a relatively low-mass stellar object followed by accretion of debris from that star (e.g., observed in 2024--2025). 
In this context,  it is important to note that specific spectral changes 
during QPE events. 
We found that as the QPE outburst evolves, the spectrum of RXJ1301 shows an increase in the soft excess at the outburst maximum, accompanied by a decrease in the Comptonization fraction ($0.1<f<0.3$), some increase in the disk photon temperature from 30 eV to 200 eV, which is consistent with the results by  
G24, who analyzed the X-ray emission of RXJ1301 using slightly different spectral models. 
However, there are discrepancies in the best-fit values for the photon index: $\Gamma=1.9-3.8$ by G24 
in contrast to our values $\Gamma=1-2.7$. We attribute these differences because of use of   different spectral models for the data interpretation.  Specifically, G24
used the {\tt Diskbb + Bbody}, {\tt Diskbb + Powerlaw}, {\tt Diskbb + Bbody + Powerlaw}, and {\tt Diskbb + CompTT} models, which differs from our spectral model, {\tt Tbabs*BMC}. 

\subsection{On the applicability of the Comptonization model BMC for the BH mass estimate in RXJ1301 using the scaling technique  \label{discussion_BMCmodel}}


The consistency of BH masses obtained by various methods (Sect.~\ref{discussion_BHmass}), as well as the established $\Gamma-N$ correlation (Fig.~\ref{saturation_all}), indicate the applicability of the scaling method using the Comptonization model {\tt BMC}. As mentioned above, the choice of a particular Comptonization model, {\tt BMC} or {\tt CompTT}, can sometimes be related to the disk or spherical accretion hypothesis, respectively, when interpreting X-ray emission of AGNs. However, the {\tt BMC} and {\tt CompTT} models are similar in that they similarly describe the Comptonization of soft (blackbody) photons on plasma electrons, in which photons acquire energy, but differ in the electron velocity and the set of parameters. The main differences lie in the contribution of the matter bulk motion and the geometry of the scattering region. The {\tt BMC} model accounts for both thermal and dynamical (converging flow) electron velocities, while the {\tt CompTT} model describes purely thermal Comptonization in a hot plasma cloud. However, since the spectrum of RXJ1301 does not extend above 2 keV, taking the latter aspect into account is not essential.

Other aspect of applicability of the Comptonization model {\tt BMC} for the BH mass estimate  using scaling is the observed positive correlation between the photon index $\Gamma$ and the mass accretion rate $\dot M$, corresponding to the spectral evolution ``the spectrum is softer when the object is brighter''. This behavior is observed in many X-ray binaries and some AGNs, such as Mrk~766 \citep{GGT14}, 
 NGC~4051 \citep{SCT18}, SDSS~J075217.84+193542.2 \citep{TS24}, HLX--1 ESO 243--49 \citep{TS16}, M101 ULX--1 \citep{tsei16b}, NGC~1566, NGC~767, Mrk~3, 1H~0707--495 \citep{TSM25} and OJ~287 \citep{TSC23}. However, XMM-Newton observations of RXJ1301 showed a ``harder spectrum when the object is brighter'' episode \citep{Sun13,Giustini20,Giustini24}, which is also confirmed in our Fig.~\ref{sp_compar_xmm}). This situation occurs when the Compton cloud is still extended, and then the photons from the Compton cloud dominate the resulting spectrum. In fact, for other missions ({\it ROSAT, ASCA, NICER}), we observe a typical spectral evolution, when the output spectrum becomes softer with increasing source luminosity (see Fig.~\ref{scaling_rx}). Of course, the origin of Comptonization (in the warm spherical corona or in the inner parts of the accretion disk) of X-rays in RXJ1301 is debatable.
On the one hand, RXJ1301 does not have hard photons above 2 keV, unlike most AGN. On the other hand, RXJ1301 resembles a very soft state of X-ray binaries. Comparison of two models, {\tt BMC} and {\tt CompTT+BB}, applied to the spectral data of RXJ1301 did not provide much clarity due to their equivalence in terms of the $\chi^2$-criterion. Future observations, in particular spectropolarimetric observations, may provide final clarification on this issue. 
In particular, a recent analysis of IXPE observations for a number of X-ray binaries and AGN has shown that the inner region of the disk, where X-ray emission is formed due to Comptonization of soft seed photons from the disk, has a flat geometry \citep{TSM25,TSS25}. 


\subsection{On the inclusion of eruptive-state spectra in the scaling relation 
\label{discussion_ES_data}}

Another aspect of application of this method to BH mass estimating in RXJ1301 using eruptive spectra is the key assumption of the method that different reference sources exhibit the same $\Gamma - N_{BMC}$ correlation shape, and the only difference is the ratio of the BH mass to the square of the distance, namely, the $M_{BH}/d^2$ coefficient. In Fig.~\ref{scaling_rx}, the black horizontal arrow highlights that the correlations for a pair of sources (e.g., the grey line of RXJ1301 and the black line of 1H~J0707--495) are self-similar, and the only difference is the BMC normalization due to the different $M_{BH}/d^2$ ratios. Thus, to obtain the BH mass in RXJ1301, the corresponding correlation of the reference source must be shifted along the $N_{BMC}$ axis to the correlation of the target sources (see Fig.~\ref{scaling_rx}). Therefore, we also tested the performance of the scaling method when applied only to the QS data.  We slightly modified the scaling method by choosing the $\Gamma\sim 2$ level as a reference point for determining $N_t$ and $N_r$ for all scaled sources on the $\Gamma-N$ diagram. Then $s_N=N_r/N_t$ (see Eq.~(\ref{mass_target1})) determined for the $\Gamma=2$ level naturally coincides with $s_N=N_r/N_t$ determined by the saturation level $\Gamma=2.7$ due to the parallelism of the $\Gamma-N$ correlation transfer along the horizontal Normalization axis. Consequently, the scaling technique provides a target BH mass value $m_t$ using any portion of the $\Gamma - N$ correlation, including both the ES and QS data in RXJ1301. Thus, the mass determination result when using only the QS data is consistent with the result of the full-sample result, which provides convincing evidence that the BH mass measurement is reliable and does not depend on the inclusion of the physically uncertain ES data.

\subsection{Comparison of scaling, timing, and other estimates of the BH mass in RXJ1301 
\label{discussion_BHmass}}

We applied the scaling technique (ST09) for the first time to the analysis of the X-ray emission from an AGN with QPE outbursts, RXJ1301, and determined the mass of the BH at its center. 
To 
further valideted 
the inferred mass of the BH in RXJ1301, we also applied a timing analysis based on results of \cite{Song20} for the XMM-Newton/EPIC-PN data 
(Fig.~1 in \cite{Song20}), when RXJ1301 was quiescent (MJD 51888.8--51889.6, $X1$).
For this observation they 
revealed the QPO signal within a frequency range of $\times 10^{-4}-8\times 10^{-4}$ Hz. 
We also fit the energy spectra to this XMM-Newton 
data with the {\tt CompTB} model to estimate the plasma temperature $T_e$ of the Compton Cloud (CC). 
Then, knowing the plasma temperature $T_e$ we can estimate the plasma velocity as
\begin{equation}
V_{plas}\sim (2kT_{e}/m_p)^{0.5}.
\label{velocity_plasma}
\end{equation} 

Using the QPO, $\nu_{peak}$ observed at about $\nu_{peak}\sim 10^{-4}$ Hz and the plasma velocity  about  $6\times10^7$  cm s$^{-1}$  for $kT_e=12$ keV (the typical plasma temperature for the CC, see also ST09)  we can estimate the size of the CC emitting the emerging spectrum. 
In fact, when analyzing physical events related to transport and time-of-flight processes in plasma, the time-of-flight $t$ of electrons through the plasma cloud is determined by the fundamental equation $t = {L}/{v_{plas}}$. The oscillation frequency $\nu$ is the reciprocal of the time of flight ($\nu = {1}/{t} = {v_{plas}}/{L}$), where $\nu$ is the oscillation frequency, $v_{plas}$ is the electron velocity, and $L$ is the characteristic size (extension) of the plasma cloud. 
Then, knowing the plasma velocity $V_{plas}$, determined by protons (see Eq.~(\ref{velocity_plasma})), and considering that electrons and protons are tightly coupled by the electron-magnetic coupling, we can state that 
the QPO frequency is determined as a ratio:
 \begin{equation}
\nu_{peak}\sim V_{plas}/L_{CC},
\label{plat_frequency}
\end{equation} 
where $L_{CC}$ is the CC size and $V_{plas}=6\times10^7$ cm~s$^{-1}$ is a typical plasma (proton) velocity  in the CC  related to the plasma temperature  of order of 12 keV. 
 From Eq.~(\ref{plat_frequency}) we estimate the CC size ($L_{CC}$) using 
 the frequency $\nu_{peak}\sim10^{-4}$ Hz in order to estimate the CC size and finally we have 

\begin{equation}
L_{CC}\sim 2 \times10^{11}\frac{V_{plas}/6\times10^7 \rm {~cm~s^{-1}}}{\nu_{peak}/3\times10^{-4}~{\rm Hz}} {\rm ~~cm}.
\label{CC_size}
\end{equation}

We use this CC size  $L_{CC}$ 
 to estimate the Schwarzschild radius of RXJ1301 
 (and then 
 the appropriate  BH mass in RXJ1301) 
 by comparing this size with the Schwarzschild radius for the Sun $R_{{\rm S}\odot}$ (recall that the Sun has a mass of 1~M$_{\odot}$ and a Schwarzschild radius $R_{{\rm S}\odot}= 2.95\times 10^5$ cm). Thus, a BH mass of RXJ1301: 
\begin{equation}
M_{1301}\sim 1~M_{\odot}\times L_{CC}/R_{{\rm S}\odot} \sim  
10^6{\rm ~~solar ~masses},  
\label{BH_estimate}
\end{equation}
\noindent which is consistent with 
the BH mass  value using the $\Gamma-\dot M$ correlation [see  Eq. (\ref{arithm_mean}), 
Table \ref{tab:par_scal_rx} and  Fig. \ref{scaling_rx}]. 

It should be noted that the available data on RX J1301 are too poor to distinguish the mode of the QPO \citep[e.g.,][]{van der Klis89, Casella05, Remillard+McClintock06} discovered by \cite{Song20} and, accordingly, the specific physical mechanism by which this QPO arises in RX1301. Therefore, our interpretation is one of the possible ones and does not claim to be definitive. 
Nevertheless, this mass, M$_{1301}\sim 10^6$ M$_{\odot}$, is consistent with our spectral scaling estimate of the BH mass in RXJ1301, $M_{1301}\sim 4\times 10^5 - 7\times 10^6)$ M$_{\odot}$ for $i=0-70^{\circ}$, and
is qualitatively consistent with BH mass estimates obtained by other methods.
%
For example, to estimate the mass of the central SMBH in RXJ1301, \cite{Sun13} used the empirical relation $M_{BH} - \sigma$ from the the width of the [O III] line and obtained the BH mass $M_{1301}^{Sun} = 8\times10^{5}$ M$_{\odot}$. Furthermore, \cite{Shu17} applied a spectral fit to the UV-to-X-ray data and estimated the BH mass in RXJ1301 $M_{1301}^{Shu} = (1.7 - 2.8)\times  10^{6}$ M$_{\odot}$. In turn, \cite{Wevers22,Wevers24} estimated the BH mass in RXJ1301 from velocity dispersion measurements and obtained $M_{1301}^{Wevers} = (1.2 - 4.5)\times 10^{6}$ M$_{\odot}$. 
Finally, \cite{Franchini23} performed a timing analysis and simulation of X-ray emission from RXJ1301 in an EMRI scenario based on a primaryl BH mass ranging from $\approx 1 \times 10^{5}$  M$_{\odot}$ to 4.5 $\times 10^6$  M$_{\odot}$. Furtermore, \cite{Franchini23} use this relatively low BH mass to simulate a distant eccentric orbital object ($e \sim 0.05 - 0.5$), where a rotating secondary star (or stellar-mass object) collides with the BH's accretion disk, causing soft X-ray outbursts  that are consistent with the observations discussed. 
Thus, the BH masses in RXJ1301 estimated by different methods are consistent with each other and range from approximately $2\times10^{5}$ to $6\times10^{7}$ M$_{\odot}$, indicating the internal validity of these methods and the reliability of the obtained mass estimate.

\subsection{On the softness of the X-ray spectrum in RXJ1301 \label{discussion_soft_spectrum}}

It is interesting to discuss the question of why the outgoing spectrum of RXJ1301 is so soft and limited to photon energies of 2 keV. This is easily interpreted in any Comptonization model ({\tt BMC} and {\tt CompTT}) and is related to the soft photon spectrum  related to the BB spectrum of the disk around a SMBH with a mass of approximately 10$^6$ M$_{\odot}$. Namely, the {\tt BMC}/{\tt CompTT} model is a convolution of the BB and the Green's Comptonization function. In particular, the temperature of the {\tt BB} photons (heated and coming from the accretion disk) is a self-consistent parameter in the {\tt BMC}model, which ultimately yields physically reasonable spectral index values for RXJ1301. 
The temperature of the seed photons in the disk precisely corresponds to the inferred mass of the BH at the accretion disk center. In fact, the peak temperature of the disk is $kT_s \le 1~{\rm keV}/(M_{BH}/10~M_{\odot})^{1/4}\sim$ 10 eV taking into account the color factor [where the BH mass $M_{BH}\sim T_{s}^{-4}$ is $\sim 10^6$ M$_{\odot}$, \cite{SS73}].  Therefore, the comparison of this temperature with the typical CC seed photon temperature is more preferable than with the ones in corona, which is located at some distance from the disk.

It should be noted, however, that RXJ1301 belongs to a rare class of ultra-soft AGNs, and its photon index -- sometimes reaching $\Gamma\sim 7$ with simple approximations of the soft range in the softest states \citep{Sun13} -- is extreme even for this rare class. While the high BH mass in RXJ1301 is consistent with why the accretion disk produces soft photons, it doesn't shed light on why other AGNs with similar BH masses don't exhibit comparable soft spectra. 
For example, the ultra-soft AGN source 1H~0707--495, which also has a BH mass of $10^6-10^7$ M$_{\odot}$, 
also exhibits a soft X-ray excess below 1 keV ($kT_s \sim 120$ eV); however, its spectrum also contains a small contribution of relatively hard X-ray emission (> 2 keV) with a photon index $\Gamma$ varying between 2 and 3 during X-ray source outbursts \citep{TSM25}. 
Note also that in RXJ1301, 1H~0707--495, and similar sources (RE~J1034+396, IRAS~13224--3809), the ultra-soft spectrum structure is continuously preserved, regardless of the outburst/quiescent state. 
In contrast, in some ultra-soft sources (e.g., RX~J0039.2--5117, IC~3599, and GSN~069), the ultra-soft spectrum appears only temporarily, during X-ray outbursts \citep{Song20}. Thus, compared to classical ultra-soft AGNs, RXJ1301 is unique in terms of the constant, extreme softness of its X-ray spectrum.

\subsection{On a possible interpretation of quisescent--eruption state evolution observed in RXJ1301 \label{discussion_QPE-interpretation}}

First of all, it should be acknowledged that the nature of QPE is still widely debated \citep{Linial+Metzger23, Franchini23,  Pan23,Shu17}. The ES emission may originate either from the BH vicinity (in which case the scaling relation could apply) or from regions far from the BH (e.g., star-disk collisions). 
The obtained results (Sect.~\ref{results}) 
allow us to  interpret the X-ray variability of RXJ1301 in an EMRI model \citep{Xian21,Franchini23},  which posits that a low-mass 
BH (or a stellar-mass object) orbits a central, more massive (primary) BH. Furthermore, the low-mass object can undergo partial disruption (in the case of a stellar-mass object), and different low-mass objects can participate at different epochs. In the case of a low-mass (secondary) BH orbiting a more massive BH, a newborn temporary disk forms around the secondary BH. The passage of the secondary BH through the accretion disk of the massive BH is accompanied by a gravitational shock and dynamical friction, which can cause the X-ray outburst. 
Then the QPE 
time is defined as $t_{accr}\sim R/V \sim 2\times 10^{11}$cm$/10^8$ cm/s $\sim$ 2~ks (i.e., exactly what is observed). 
Note that our interpretation within the EMRI model 
is one of several possible interpretations and certainly does not claim to be definitive. 
In particular, within the EMRI scenario for QPE \citep{Linial+Metzger23, Franchini23}, ES emission originate at the star-disk collision region, distant from the BH. The size and properties of the ejection region in this case may not be directly related to the BH mass.
 Alternatively, prolonged post-outburst emission could also occur when X-ray emission is fueled by the cooling of the shock-heated ejecta \citep{Linial+Metzger23,Bychkov+Seifina96} 
under the assumption of accretion disk instability \citep[][e.g.,]{Shu17,Pan23}. Under such a model, the ES emission could indeed originate from the inner disk, and thus the scaling relation might in principle apply.
However, the emission time in this case is much shorter (approximately 1 $\mu s$) than in the case of disk accretion (1--2 ks, see Fig.~3 in \cite{Sun13} and Fig.~1 in \cite{Giustini24}). 


This scenario is also consistent with a lack of correlation of radio/X-ray variability during QPEs in RXJ1301 (see G24), as the X-ray and radio emission regions are widely separated.
In this case, the formation of the  radio jet is not initiated by an accretion rate enhancement due to QPEs from orbital disk punctures, when the most of the stellar-mass object's  
disk debris is captured by the SMBH rather than ejected as an episodic radio jet.

Given that the physical origin of the QPE remains uncertain, applying the scaling method to the RXJ1301 data can be considered as one possible way to estimate the central BH mass, consistent with the results of other authors using other methods. Furthermore, as we demonstrated previously, the scaling method applied to the QS data yields the same BH mass as the full-sampling (QS and ES) case. This proves that the BH mass measurement in RXJ1301 is reliable using the scaling technique. 
\section{Conclusions \label{summary}} 

We present our analysis  of the spectral properties observed in X-rays from  RXJ1301 during QPE events. We analyze the source observations by  ASCA, {\it NICER, XMM-Newton}  and {\it ROSAT}. We show that the X-ray spectra for RXJ1301 during all spectral states can be well fitted by 
the  Comptonization model. 

We detected new outburst episodes in RXJ1301 in 2024 in the {\it NICER} light curves, accompanied by a softening of the spectrum during the outbursts.  In the QS, the spectrum appears to be dominated by a Comptonized component, the temperature of whose seed photons is comparable to the predicted temperature of the accretion disk. In particular, we  found that  the QPE outburst led to an increase in the soft excess  of  the RXJ1301  spectrum  accompanied by a decrease in the illumination fraction (which varies in the range $0.1<f<1$) 
and  some increase of the disk photon temperature $kT_s$ from 20 eV to 200 eV.

 We discovered the saturation of the spectral index for RXJ1301 during QPE outbursts  (at $\Gamma_{1301}=2.7$) and fist determined  the mass of the SMBH in the center of this AGN applying  the scaling technique: 
 from $4\times 10^5$ M$_ {\odot}$ to  $7\times 10^6$ M$_ {\odot}$ for the inclination from 0$^{\circ}$ to 70$^{\circ}$, 
 by the scaling method. Our obtained SMBH mass is consistent with that of RXJ1301 estimated using the empirical $M_{BH} - \sigma$ relation \citep{Sun13}, using spectral fitting to UV and X-ray data \citep{Shu17}, and based on velocity dispersion measurements \citep{Wevers22,Wevers24}, indicating the reliability of our results.

\begin{acknowledgements}
We appreciate valuable remarks  by Chris Shrader on the paper.  We would like to thank Juliana Seifina for her assistance in preparing  the figures.  
 We also acknowledge the reviewers and editor for their in-depth analysis of the paper 
and useful comments that improved our paper. The study was conducted under the state assignment of Lomonosov Moscow State University.
\end{acknowledgements}


\end{document}